\documentclass[12pt]{article}

\usepackage{amsthm,amsmath,bbm,amssymb}
\usepackage[status=draft,inline,index]{fixme}
\usepackage{graphicx,subcaption}
\usepackage[margin=1.25in]{geometry}
\usepackage[square,numbers]{natbib}
\usepackage[normalem]{ulem}
\newtheorem{assumption}{Assumption}

\newtheorem{proposition}{Proposition}

\newcommand{\independent}{\perp\mkern-9.5mu\perp}

\usepackage{xr} % Reference between .tex files
\usepackage{xr} % Reference between .tex files

\makeatletter
\newcommand*{\addFileDependency}[1]{% argument=file name and extension
  \typeout{(#1)}
  \@addtofilelist{#1}
  \IfFileExists{#1}{}{\typeout{No file #1.}}
}
\makeatother

\newcommand*{\myexternaldocument}[1]{%
    \externaldocument{#1}%
    \addFileDependency{#1.tex}%
    \addFileDependency{#1.aux}%
}

\myexternaldocument{appendix}
\allowdisplaybreaks % Allow page breaks in align environment

\usepackage{setspace} 

\usepackage{tikz}
\usetikzlibrary{arrows.meta,shapes}
\usetikzlibrary{arrows,shapes.arrows,shapes.geometric,shapes.multipart,
decorations.pathmorphing,positioning,shapes.swigs,}

\fxsetup{theme=color,margin=false,mode=multiuser}
\FXRegisterAuthor{uo}{uoe}{\color{orange}UO}
\FXRegisterAuthor{ms}{mse}{\color{blue}MS}
\FXRegisterAuthor{dn}{dne}{\color{magenta}DN}
\title{Distinguishing immunological and behavioral effects of vaccination}

\author{
Mats Stensrud$^{1}$, Daniel Nevo$^{2}$, Uri Obolski$^{3,4}$ \\
{{\small $^{1}$Department of Mathematics, \'Ecole Polytechnique F\'ed\'erale de Lausanne, Switzerland} } \\
{{\small $^{2}$Department of Statistics and Operations Research, Tel Aviv University, Israel
 }} \\
 {{\small $^{3}$Department of Epidemiology and Preventive Medicine, Tel Aviv University, Israel }} \\
 {{\small $^{4}$Department of Environmental Studies, Tel Aviv University, Israel }}
}

\date{}
\begin{document}
\maketitle

\begin{abstract}
The interpretation of vaccine efficacy estimands is subtle, even in randomized trials designed to quantify immunological effects of vaccination. In this article, we introduce terminology to distinguish between different vaccine efficacy estimands and clarify their interpretations. This allows us to explicitly consider immunological and behavioural effects of vaccination, and establish that policy-relevant estimands can differ substantially from those commonly reported in vaccine trials. We further show that a conventional vaccine trial allows identification and estimation of different vaccine estimands under plausible conditions, if one additional post-treatment variable is measured. Specifically, we utilize a ``belief variable'' that indicates the treatment an individual believed they had received. The belief variable is similar to ``blinding assessment'' variables that are occasionally collected in placebo-controlled trials in other fields. We illustrate the relations between the different estimands, and their practical relevance, in numerical examples based on an influenza vaccine trial. 
\end{abstract}


\section{Introduction}
\label{sec: intro}

Pharmaceutical interventions are commonly justified by their immunological mechanisms of action, but  might also affect outcomes through other pathways. For example, recipients of vaccines have been reported to increase their number of social contacts due to perceived protective effects \citep{serisier2023case, brewer2007risk, hossain2022scaling}, although the extent of such behavioral changes varies across populations and time \citep{goldszmidt2021protective, wright2022people,hall2022vaccinated,thorpe2022self}.

Conventional vaccine trials are designed to identify \textit{immunological} effects of vaccines \citep{rid2014placebo}. These trials often have blinded treatment and control groups \citep{bender2022really,haas2022frequency} and the rationale for (patient) blinding is precisely to eliminate non-immunological effects of vaccination. Indeed, an ideal placebo control satisfies two criteria: it does not have any cross-reactivity with the pathogen in question; and it is perceived to be indistinguishable from the vaccine, for example by inducing common vaccine side effects like fever or soreness in the place of injection. The second criterion is challenging to satisfy in many vaccine trials where inert saline vaccines are used as controls, see e.g.\ Haas et al \cite{haas2022frequency}.

The reliability of placebo controls has been studied in so-called unblinding assessments or manipulation checks \citep{moustgaard2020impact,ejelov2020rarely}, where trial participants are asked to guess the treatment they received. Differences in guesses between assignment groups indicate that the placebo was unsuccessful. Such differences have been observed in trials assessing appetite suppressive treatments \citep{moscucci1987blinding}, smoking cessation strategies \citep{schnoll2008can}, psychiatric drugs \citep{fisher1993sound, scott2022systematic}, back pain treatments \citep{freed2021blinding} and other interventions \citep{boutron2005review, hrobjartsson2007blinded}.

However, unblinding might be a consequence of the treatment being effective. If a treatment is noticeably beneficial, and individuals are asked to guess their treatment group after the effect becomes evident, then these individuals might correctly guess their treatment status. Because unblinding assessments are difficult, the mandatory reporting of blinding success was revoked in the 2001 CONSORT guidelines \citep{polit2011deliberate,webster2021measuring}. Thereafter, assessment of blinding in RCTs has become less frequent \citep{boutron2005review, hrobjartsson2007blinded,scott2022systematic, bello2014risk,kahan2015blinded}. In particular, we could not find examples of blinding assessment in vaccine studies. 

In this article we claim that an assessment of treatment beliefs, similar to unblinding assessments, is desirable in vaccine trials because this assessment can allow us to identify policy-relevant estimands. First, we formally study causal effects targeted by vaccine trials, scrutinizing their practical relevance. Even under perfect blinding and no interference, the conventional vaccine efficacy estimated from a trial might not be representative of real-world vaccine effectiveness. In addition, broken blinding challenges the interpretation of common vaccine estimands as parameters quantifying ``immunological efficacy''.
 Second, we describe how different, but related, estimands can be identified and estimated from conventional vaccine trials under testable, and often plausible, assumptions when a blinding assessment is conducted.

\section{Preliminaries}
\label{sec: prelim}
 
Consider data from a \textit{blinded} randomized controlled trial (RCT) with $n$ individuals who are assigned treatment $A \in \{0,1\}$ at baseline, where $A=1$ indicates receiving vaccine and $A=0$ indicates placebo or other control.  Together with $A$, we explicitly define the \textit{message} $M \in \{-1,0,1\}$, indicating whether the individual receives the \textit{message} that the vaccine status is blinded ($M=-1$), that they are unvaccinated $(M=0)$ or that they are vaccinated  against
the pathogen of interest $(M=1)$. Unlike a real-world setting, where we would expect that $A=M$ with probability one (w.p.1), blinding in an RCT implies that the message $M$ is fixed to $-1$. %An analogous \textit{decomposition of the intervention} into $M$ and $A$ has previously been suggested by \citet{didelez2018defining} to motivate interventionist mediation estimands, see also \citet{robins2010alternative,stensrud2019separable,robins2020interventionist}. 

Furthermore, let $L$ be a vector of measured covariates, which might affect the outcome $Y$.  We treat $L$ as discrete for simplicity of presentation, but all results hold for continuous $L$ by replacing probabilities with probability density functions and sums with appropriate integrals. Let $S \in \{0,1\}$ be an indicator of a possible side effect, and let $B \in [0,1]$ be a variable quantifying the degree to which an individual \textit{believes} that they have received the vaccine, where $B=1$ corresponds to being convinced about having received the active vaccine and $B=0$ being convinced about having received control; here we would expect that the belief depends on the type of control, for example depending on whether the control is simply no treatment or an inert (placebo) vaccine. Finally, let $E$ be a variable quantifying how much an individual has been exposed to the infectious agent. We do not assume that $E$ is measured and leave the domain of $E$ arbitrary. In addition, we will occasionally introduce an unmeasured variable $U$ as a common cause of at least two of previously introduced variables.

As assumed in most analyses of vaccine RCTs, suppose that the trial participants are drawn from a near-infinite super-population where interactions among participants are negligible. Therefore, we omit the $i$ subscript from random variables. The assumption that interactions between participants in the trial and the population are negligible with respect to the vaccine effect implies an assumption about no interference, so all potential outcomes we subsequently present are well defined. However, our arguments also apply to certain situations when interference is present, as discussed in \ref{sec: no interference}.

We use superscripts to denote counterfactuals. In particular, let $B^{a,m}$ be an individual's belief about their vaccine status when treatment and the message are set to $a$ and $m$, respectively. When there is no blinding, e.g. after the vaccine has been made available for the entire population and $A=M$ w.p.1, we would expect that $B^{a,m}=a=m$. Let also $E^{a,m}$ quantify the exposure of the study participant when treatment is fixed to $a$ and the message to $m$. As with $E$, the domain of $E^{a,m}$ is left arbitrary. Hence, our results do not depend on the variable type (e.g. binary, count or continuous) assumed for $E^{a,m}$. If receiving the vaccine can cause a side effect shortly after vaccination, say within 7 days \cite{haas2022frequency}, we further define $S^a$ to be the indicator that the participant experienced this side effect. Let $Y^a$ be the disease status some fixed time (e.g. one month) after an individual was assigned treatment level $A=a$, which is measured without misclassification. Finally, let $Y^{a,m}$ be the outcome had  treatment level was fixed to $A=a$ and the message to $M=m$ . Henceforth we assume consistency with respect to all counterfactuals defined and corresponding observed data, for example, if $A=a$ then $Y=Y^a$.

\section{Causal effects and target trials}
\label{sec: causal effects}

\subsection{Conventional two-arm trial}
\label{sec: causal effects two-arm}

Consider first the average treatment effect of being vaccinated ($A$) on the clinical outcome ($Y$), when the treatment allocation is fixed to be blinded $(m=-1)$, 
\begin{align}
& \mathbb{E} ( Y^{a=1,m=-1}) \text{ vs. }  \mathbb{E} ( Y^{a=0,m=-1}).
\label{eq: ate}
\end{align}
This effect is identified by design in a conventional, blinded two-arm vaccine trial, henceforth denoted by $\mathcal{T}_{II}$. We deliberately set $m=-1$ as part of the intervention indicated in the superscripts, because blinding is a crucial feature of the intervention tested in $\mathcal{T}_{II}$. While studies of vaccine effects usually state estimands of the form $\mathbb{E} ( Y^{a=1})$ and $\mathbb{E} ( Y^{a=0})$, without indicating the message $M$, we make this distinction to clarify that other, subtly different estimands can be important for policy decisions. Our variable definitions are related to, but different from, the definitions given by Murray \cite{murray2021demystifying}, who explicitly formulated a counterfactual definition of a per-protocol placebo effect, see \ref{sec: per protocol murray}.

Because conventional vaccine trials enforce $m=-1$, such trials are, at least implicitly, targeting an immunological vaccine effect: the intention of blinding is to eliminate certain (psychological and behavioral) effects of receiving the vaccine. Suppose now that we offer the vaccine to individuals in a real-world setting, outside the trial. Even if the trial and real-world settings share similar conditions, e.g.\ individuals are drawn from the same super-population, and the transmission rates are equal in both settings, the effect of the vaccine in the real-world setting might differ from the effect in the RCT. Individuals in the real-world setting are,  unlike the vaccinated trial participants, informed about the treatment they received ($M=A$ w.p.1). In particular, a vaccinated individual \textit{knows} that they received a vaccine ($B=1$) and this knowledge can lead to changes in their behavior. For example, the vaccinated individual might reduce their protective behaviors, and thus increase their risk of being exposed. %Figure \ref{fig:vac obs study} presents a DAG describing such a non-experimental setting. In the figure, the bold arrow from $A$ to $M$ indicates vaccination status is known and  the arrow from $M$ into $E$ indicates  the effect of knowledge of vaccination status on exposure status.

Because people might change behavior after vaccination, the \textit{total effect},
\begin{align}
    \mathbb{E}(Y^{a=1,m=1}) \text{ v.s. } 
    \mathbb{E}(Y^{a=0,m=0}) \label{eq: total eff},
\end{align}
which quantifies the joint effect of receiving both the vaccine ingredient and the message, is a relevant parameter for policy makers when deciding vaccine policies. This effect
is different from the the effect given in \eqref{eq: ate}. That is, the effect in \eqref{eq: ate} is analogous to the effect targeted in a successfully blinded experiment, where the intention might be to eliminate placebo effects by fixing $m=-1$; the effect in \eqref{eq: total eff} is the effect in an unblinded trial, and captures different pathways by which vaccination can affect the outcome of interest. For example, if vaccination leads to reduced use of protective measures, the knowledge of being vaccinated might counteract the protective immunological  effect of the vaccine.

However, the effect in \eqref{eq: total eff} is not identifiable from the data in $\mathcal{T}_{II}$ without additional assumptions. We will therefore consider hypothetical trials that allow identification of such effects, including a feasible, minor modification of $\mathcal{T}_{II}$.

\subsection{Hypothetical four-arm trial}
\label{sec: hyp four-arm trial}
 Consider a four-arm trial  where each individual is assigned the blinded treatment $A \in \{0,1\}$, and then immediately given a message $M \in \{0,1\}$, stating that they, possibly contrary to fact, received the control ($M=0$) or active vaccine ($M=1)$. In this trial, henceforth denoted by $\mathcal{T}_{IV}$, it is possible that the \textit{message} ($M$) is the opposite of the actual treatment assignment ($A$), that is, $\Pr(M \neq A) > 0$.  
 By design, in $\mathcal{T}_{IV}$ we identify
\begin{align*}
    \mathbb{E}(Y^{a,m}) \text{ v.s. } 
    \mathbb{E}(Y^{a',m'}),
\end{align*}
for $a,a',m,m' \in \{0,1\}$.

Such a trial design, also known as a ``balanced placebo design'' \citep{colagiuri2010participant}, has been implemented
 to examine the effects of nausea treatments \citep{roscoe2010exploratory}, nicotine and  alcohol \citep{colagiuri2010participant}, and caffeine \citep{beedie2006placebo}. To the best of our knowledge, this design has never been implemented to study vaccine effects. Conducting such a vaccine trial is ethically problematic, because the participants are given misleading information about vaccination status that might e.g. affect their risk through behavior \citep{colagiuri2010participant}. Even if the trial is practically infeasible, we can still conceptualize a study that jointly assigns $A$ and $M$ at random, which would allow us to \textit{separate} immunological and behavioral effects of receiving the vaccine. For example, the contrast 
\begin{align}
    \mathbb{E}(Y^{a=1,m}) \text{ v.s. } 
    \mathbb{E}(Y^{a=0,m}) \label{eq: direct eff}
\end{align}
is, like the contrast in \eqref{eq: ate},  expected to quantify an immunological effect of receiving the vaccine, because individuals in both arms are told that they have the same vaccination status, $m \in \{0,1\}$.

On the other hand, the contrast
\begin{align}
    \mathbb{E}(Y^{a,m=1}) \text{ v.s. } 
    \mathbb{E}(Y^{a,m=0}) \label{eq: indirect eff}
\end{align}
quantifies a behavioral effect of the vaccine, in the sense that both groups receive the same biological vaccine component ($a$), but one of the groups are told that they, contrary to fact, did not. Thus, this contrast quantifies how knowledge (belief) of being vaccinated changes the outcome, e.g.\ through risk-increasing behavior. Furthermore, the total effect, \eqref{eq: total eff},  would be identified from $\mathcal{T}_{IV}$  without additional assumptions. 

Because of the ethical and logistical issues with conducting $\mathcal{T}_{IV}$, we  present conditions ensuring that we can use data from $\mathcal{T}_{II}$ to identify and interpret \eqref{eq: direct eff} and \eqref{eq: indirect eff} as immunological and behavioral effects, respectively. %These results are important, because the four-arm trial is rarely, if ever, feasible to implement in practice. 

\subsection{Identification based on relations between the two-arm and four-arm trials}
\label{sec: identification}
To relate the outcomes in $\mathcal{T}_{II}$ and $\mathcal{T}_{IV}$, consider the belief $B$, quantifying the degree to which an individual believes that they received active vaccine (higher values of $B$) or control (lower values of $B$). In particular, $B=1$ means that the individual is convinced that they were vaccinated. While the results we present in this work are applicable for continuous $B \in [0,1]$, to simplify the notation we henceforth focus on a binary belief, $B\in \{0,1\}$.

If the four-arm trial $\mathcal{T}_{IV}$ is successful, we would expect that the message $M$ deterministically causes the belief $B$.
 \begin{assumption}
 $B = M $ w.p.1. where $M \in \{0,1\}$.
 \label{ass: belief determinism}
 \end{assumption}

In $\mathcal{T}_{II}$, individuals receive no message, which we defined as fixing $m=-1$, but they might still form beliefs about the treatment received. When the belief $B$ affects the risk of exposure $E$, the counterfactual quantity identified in $\mathcal{T}_{II}$, $\mathbb{E}(Y^{a=1,m=-1})$, would be less relevant to the outcomes in the  setting where people know their vaccination status, as is usually the case in practice. 
 
However, it is feasible to measure the belief $B$ in $\mathcal{T}_{II}$,  by asking whether an individual believes they received active vaccine or placebo. We denote by $\mathcal{T}_{II^B}$ the two-arm trial where $B$ is also measured. By introducing the belief variable $B$, we can formalize the notion that receiving the vaccine affects our risk of infectious disease outcomes both through the immunological effect of a vaccine and through behavior. 

Suppose first that the belief determinism holds in the four-arm trial $\mathcal{T}_{IV}$, that is $B = M$ w.p.1. Consider now the six-arm trial which incorporates the arms of the two- and four-arm trials introduced so far: let $\mathcal{T}_{VI}$ be the trial where $A \in \{0,1\}$  and $M
 \in \{-1,0,1\}$ are randomly assigned jointly, but independently of each other.  Suppose further that the message $M$ only affects $Y$ through the belief, which we explicitly state as an isolation condition. 
\begin{assumption}[$M$ partial isolation]
\label{ass: m partial isolation}
The only causal paths from $M$ to $Y$ are directed paths intersected by $B$.
\end{assumption}
In our setting, Assumption \ref{ass: m partial isolation} requires that the external message about the treatment status only affects the outcome $Y$ through our belief about the treatment status. Assumption \ref{ass: m partial isolation} seems to be plausible in practice; to violate this assumption, the message $M$ must affect $Y$ outside of the belief $B$, which will by contrived in many settings. For example,  Assumption \ref{ass: m partial isolation} holds in the DAGs in Figure \ref{fig: dag two-arm trial}, where the arrow from $M=-1$ to $B$ is trivial, and in Figure \ref{fig: dag four-arm trial}.

Consider also the following assumption, inspired by previous work on separable effects \cite{robins2010alternative,stensrud2019generalized,stensrud2022conditional}.

\begin{assumption}[$Y$ Dismissible component condition]
\label{ass: delta 1}
$$
%Y(G) \independent_{VI} M(G) \mid B(G), L(G),A(G) ,
Y \independent_{\text{VI}} M \mid L, A, B,
$$ 
where  $\independent_{\text{VI}}$ denotes independence in $\mathcal{T}_{VI}$. 
\end{assumption}

We can directly read off that this assumption holds with $L = \emptyset$ in the DAG in Figure \ref{fig: dag six-arm trial}, describing a six-arm trial $\mathcal{T}_{VI}$. In Figure \ref{fig: dag six-arm trial}, the node $E$ is not needed to evaluate our identification conditions, but including $E$ clarifies that $M$ only affects $Y$ through the exposure to the infectious agent, $E$.

\begin{figure}
    \begin{minipage}{1\textwidth}
        \centering
                \begin{tikzpicture}
                    \tikzset{line width=1.5pt, outer sep=0pt,
                    ell/.style={draw,fill=white, inner sep=2pt,
                    line width=1.5pt},
                    swig vsplit={gap=5pt,
                    inner line width right=0.5pt}};

                    \node[name=A,ell,  shape=ellipse] at (-6,-1) {$A$};
                    \node[name=M,ell,  shape=ellipse] at (-6,1) {$M=-1$};
                    \node[name=B,ell,  shape=ellipse] at (-3,0) {$B$};
                    \node[name=E,ell,  shape=ellipse] at (0,0) {$E$};
                    \node[name=U,ell,  shape=ellipse] at (0,2) {$U$};
                    \node[name=Y,ell,  shape=ellipse] at (3,0) {$Y$};
                     \begin{scope}[>={Stealth[black]},
                                  every edge/.style={draw=black,very thick}]
                        \path [->] (E) edge (Y);
                   %     \path [->] (A) edge (B);
                        \path [->] (B) edge (E);
                        \path [->] (U) edge (E);
                        \path [->] (U) edge (Y);
                        \path [->] (E) edge (Y);
                        \path [->] (M) edge  (B);
                        \path [->] (U) edge[color=blue] (B);
                        \path [->] (A) edge[bend left=-30] (Y);    
                    \end{scope}
                \end{tikzpicture}
\end{minipage}
\caption{\label{fig:vac swig unmeasured conf}DAG compatible with the two-arm trial $\mathcal{T}_{II}$ where $A \in \{0,1\}$ and $M=-1$. Here, Assumption \ref{ass: delta 1} is expected to be violated due to the blue arrow.}

\end{figure}
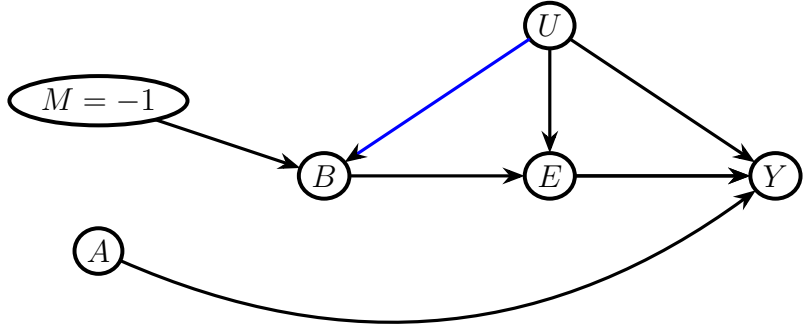

Assumption \ref{ass: delta 1} would be expected to fail, except in some special cases with perfect cancellations, whenever Assumption \ref{ass: m partial isolation} fails. However, Assumption \ref{ass: delta 1} can also fail when Assumption \ref{ass: m partial isolation} holds, e.g.\ when there are unmeasured common causes between $B$ and $Y$, as illustrated by the path  $B\leftarrow U \rightarrow Y$ in  Figure \ref{fig:vac swig unmeasured conf}. 

To identify $\mathbb{E}(Y^{a,m })$ from $\mathcal{T}_{II^B}$, we also require the following positivity assumption.
\begin{assumption}[Positivity]
\label{ass: positivity two-arm}
In $\mathcal{T}_{II}$, for all $a,b \in \{0,1\}$,  and for all possible values of $L$
\begin{equation*}
\Pr(B = b \mid L=l, A=a) > 0.
\end{equation*}
%\begin{align}
%& \Pr(L=l) > 0  \implies \nonumber\\ 
%& \Pr(B = b \mid L=l, A=a) > 0 
%\end{align}%
\end{assumption}

The positivity assumption requires that we observe individuals who believe they are vaccinated ($B=1$) and unvaccinated ($B=0$)  for each covariate value $L=l$ and treatment arm $A=a$.

The following proposition establishes that $\mathbb{E}(Y^{a,m })$ can be identifiable from $\mathcal{T}_{II^B}$. 

\begin{proposition}
\label{prop: identification simplest}
Under Assumptions \ref{ass: belief determinism}, \ref{ass: delta 1} and \ref{ass: positivity two-arm}, $\mathbb{E}(Y^{a,m}) $  for $a,m \in \{0,1\}$ is identified  from the two-arm trial $\mathcal{T}_{II^B}$,
\begin{align}
\label{eq: ident prop 1}
\mathbb{E}(Y^{a,m })  = \sum_l \mathbb{E}(Y \mid L = l, A = a, B = m)\Pr(L=l) \quad \text{ for } a,m \in \{0,1\}. 
\end{align}
\end{proposition}
See \ref{sec: proofs} for a proof.

\subsection{Decomposition into immunological and behavioral effects}
Suppose that receiving the vaccine affects the risk of being exposed ($E$) only through the message, $M$, as formalized in the following assumption.

\begin{assumption}[No direct effect of $A$ on exposure $E$]
\label{ass: no dir on exposure}
$$
E^{a=1,m} = E^{a=0,m} \text{ }w.p.1 \text{ for } m \in \{-1,0,1\}.
$$    
\end{assumption}
Assumption \ref{ass: no dir on exposure} can in principle be tested in a trial where $(A,M)$ are randomly assigned and the exposure $E$ is measured, e.g.\ by assessing whether
$$
\mathbb{E}(E \mid A=1, M=m) = 
\mathbb{E}(E \mid A=0, M=m) 
$$ for either value of $m$. Measuring $E$ is practically difficult and rarely done in trials, but augmented information on behavior could at least in principle be collected, e.g. using smartphone data \citep{pandit2022smartphone}. However, such information is often hard to obtain \citep{stensrud2022identification}. Yet Assumption \ref{ass: no dir on exposure} seems to be plausible whenever the vaccine does not induce noticeable (side) effects, e.g.\ when blinding is successful. 

When both Assumptions \ref{ass: m partial isolation} and \ref{ass: no dir on exposure} hold, we can interpret \eqref{eq: direct eff} and \eqref{eq: indirect eff} as direct and indirect effects, quantifying immunological and behavioral components, respectively. However, Assumptions \ref{ass: m partial isolation} and \ref{ass: no dir on exposure} are not necessary for any of our mathematical (identification) arguments to be valid. We further discuss decompositions of the total effect \eqref{eq: total eff} on different scales in  \ref{sec: decompositions}.

\subsection{Effects under broken blinding}
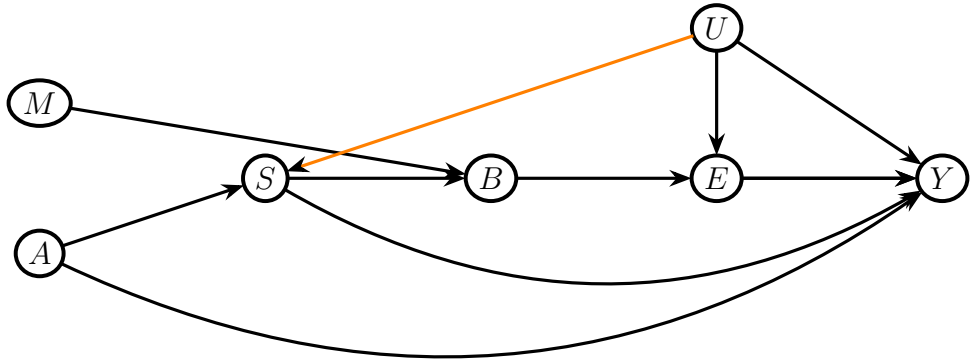
\begin{figure}
    \begin{minipage}{1\textwidth}
        \centering
                \begin{tikzpicture}
                    \tikzset{line width=1.5pt, outer sep=0pt,
                    ell/.style={draw,fill=white, inner sep=2pt,
                    line width=1.5pt},
                    swig vsplit={gap=5pt,
                    inner line width right=0.5pt}};
                    \node[name=A,ell,  shape=ellipse] at (-9,-1) {$A$};
                    \node[name=S,ell,  shape=ellipse] at (-6,0) {$S$};
                    \node[name=B,ell,  shape=ellipse] at (-3,0) {$B$};
                    \node[name=E,ell,  shape=ellipse] at (0,0) {$E$};
                    \node[name=U,ell,  shape=ellipse] at (0,2) {$U$};
                    \node[name=Y,ell,  shape=ellipse] at (3,0) {$Y$};
                    \node[name=M,ell,  shape=ellipse] at (-9,1) {$M$};
                     \begin{scope}[>={Stealth[black]},
                                  every edge/.style={draw=black,very thick}]
                        \path [->] (E) edge (Y);
                        \path [->] (A) edge (S);
                        \path [->] (S) edge (B);
                        \path [->] (M) edge (B);
               %         \path [->] (A) edge[line width=0.85mm] (M);
                        \path [->] (B) edge (E);
                        \path [->] (U) edge (E);
                        \path [->] (U) edge[color=orange] (S);
                        \path [->] (U) edge (Y);
                        \path [->] (E) edge (Y);
                        \path [->] (S) edge[bend left=-30] (Y);    
                        \path [->] (A) edge[bend left=-30] (Y);    
                    \end{scope}
                \end{tikzpicture}
\end{minipage}
\caption{\label{fig:vac sb}DAG describing the six-arm trial  $\mathcal{T}_{VI}$, where a side effect $S$ can affect the belief $B$. Here, the presence of the orange arrow violates Assumption \ref{ass: delta 1} but not Assumption \ref{ass: delta 2}.}
\end{figure}

Consider the assumption that the (blinded) treatment $A$ has no effect on the belief $B$ in $\mathcal{T}_{II}$, which we denote successful blinding. 
\begin{assumption}[Successful blinding] \label{ass: no effect on belief}
\begin{align*}
   B^{a=1,m=-1} = B^{a=0,m=-1}.
\end{align*}
\end{assumption}
Assumption \ref{ass: no effect on belief} is easily falsifiable in $\mathcal{T}_{II^B}$, e.g.\ by evaluating whether 
$$
\Pr(B=1 \mid A=1) = 
\Pr(B=1 \mid A=0 ) 
$$
in this trial. Assumption \ref{ass: no effect on belief} might fail in practice, indicating that \textit{blinding is broken}. Consider for example a mild side effect $S$, which is biologically induced by the vaccine. Thus, the side effect occurs more often under $A=1$ compared to $A=0$, and individuals who experience this side effect are more likely to believe that they are treated. A setting where $A$ affects the belief $B$ through a side effect $S$ is illustrated by the arrows $A\rightarrow S$ and $S\rightarrow B$ in Figure \ref{fig:vac sb}. Furthermore,  $S$ can be affected by unmeasured factors $U$ that also affect the outcome $Y$, which would imply that Assumption \ref{ass: delta 1} fails. Suppose, however, that two less restrictive dismissible component conditions hold in $\mathcal{T}_{VI}$:

\begin{assumption}[$Y,S$ Dismissible component conditions]
\label{ass: delta 2}
\begin{align}
%    Y(G) & \independent M(G) \mid B(G), L(G),A(G), S(G) , \label{eq: delta m b}\\
% B(G) & \independent A(G) \mid M(G), L(G),S(G).  \label{eq: delta a m}
    Y & \independent_{VI} M \mid  L, A, S, B, \nonumber%\label{eq: delta m b}
    \\
% B & \independent_{VI} A \mid M, L,S, \label{eq: delta a m} \\
 S & \independent_{VI} M \mid L, A  \label{eq: delta m s},
\end{align}
where $\independent_{\text{VI}}$ denotes  independence in $\mathcal{T}_{VI}$. 
\end{assumption}
Suppose also that the following positivity condition holds, which is testable using the observed data from   $\mathcal{T}_{II^B}$. 
\begin{assumption}[Positivity 2] In $\mathcal{T}_{II^B}$, for all  possible values of $L$
 \label{ass: positivity side eff}
%For the target parameter $\mathbb{E} (Y^{a,m})$ for $m \in \{0,1\}$, 
\begin{align*}
%& \Pr(L=l) > 0  \implies \nonumber\\ 
&  \Pr(S=s \mid L=l, A=a) > 0, \text{ for all }  a, s\in\{0,1\},% \nonumber
\\
&  \Pr(B=b \mid L=l, A=a, S=s) > 0, \text{ for all }  a, b, s\in\{0,1\}.
\end{align*}%
% \begin{align}
% & \Pr(L=l) > 0  \implies \nonumber\\ 
% & \quad \Pr(S=s \mid L=l, A=a) > 0, \text{ for all }  s, a\in\{0,1\}
% \end{align}%
% and 
% \begin{align}
% & \Pr(S=s,L=l) > 0  \implies \nonumber\\ 
% & \quad \Pr(B=b \mid L=l, S=s, A=a) > 0, \text{ for all }  a, b\in\{0,1\}
% \end{align}%
\end{assumption}

The following proposition establishes that $\mathbb{E}(Y^{a,m })$ can be identified from $\mathcal{T}_{II^B}$, even under broken blinding. 

\begin{proposition}
\label{prop: identification with s}
Under Assumptions  \ref{ass: belief determinism},  \ref{ass: delta 2} and  \ref{ass: positivity side eff}, $\mathbb{E}(Y^{a,m}) $ for $a,m \in \{0,1\}$ is identified from the two-arm trial $\mathcal{T}_{II^B}$, 
\begin{align}
\label{eq: ident formula S}
\mathbb{E}(Y^{a,m })  = \sum_{s,l} \mathbb{E}(Y \mid L=l, A=a, S=s,B=m) \Pr(S=s \mid L=l, A=a) \Pr(L=l). 
\end{align}
\end{proposition}
See \ref{sec: proofs} for a proof.

\subsection{Conditional causal effects}

We can use data from $\mathcal{T}_{II^B}$ to identify other effects of vaccination that are not affected by behavior.  Consider the contrast
\begin{align}
    \mathbb{E}(Y^{a=1,m=-1} \mid B^{a=1,m=-1} = b)  \text{ v.s. } 
    \mathbb{E}(Y^{a=0,m=-1} \mid B^{a=0,m=-1} = b) . \label{eq: cond outcome contrast}
\end{align}
This contrast is not a causal effect, in the sense that this contrast is not a comparison of counterfactual outcomes in the same subset of individuals. However, when Assumption \ref{ass: no effect on belief} holds, we can rewrite \eqref{eq: cond outcome contrast} as
\begin{align}
\label{eq: cond eff}
    \mathbb{E}(Y^{a=1,m=-1} \mid B^{m=-1}=b)  \text{ v.s. } 
    \mathbb{E}(Y^{a=0,m=-1} \mid B^{m=-1}=b) ,
\end{align}
which is a causal effect of treatment $a=1$ v.s. $a=0$ among those with a particular belief $b$ when $m=-1$. In $\mathcal{T}_{II^B}$, this causal effect is simply identified as 
\begin{align*}
    \mathbb{E}(Y \mid A = 1, B = b)  \text{ v.s. } 
    \mathbb{E}(Y \mid A = 0, B = b),
\end{align*}
which has an interpretation as an immunological effect under Assumption \ref{ass: no effect on belief}, as we condition on individuals having the same behavior, and thus the same risk of exposure to the infectious agent. It follows that ``immunological effects'' are not uniquely defined, because a different immunological effect was defined in \eqref{eq: direct eff}. However, \eqref{eq: cond eff} is restricted to a subset of the population that has a particular belief under blinding. It is not clear how the effect in this subpopulation is relevant to the entire population, without imposing additional assumptions.

 %In our identification arguments, we have considered a hypothetical six-arm trial $\mathcal{T}_{VI}$ with independent randomization of $M \in \{-1,0,1\}$ and $A \in \{0,1\}$, covering all the arms in the two- and four-arm trials considered so far. 

\section{Estimation}
\label{sec: estimation}

Based on the identification results in Section \ref{sec: identification}, we can motivate standard estimators of $\mathbb{E}(Y^{a,m}) $ for $ \ a,m \in \{0,1\}$ using data from $\mathcal{T}_{II^B}$  \citep{robins1986new,robins2000marginal}. Confidence intervals can, for example, be calculated by bootstrap or the delta method. %Here we apply well-known arguments for deriving g-formula and inverse probability weighted estimators \citep{robins1986new,robins2000marginal}. For both methods, inference can be carried out using asymptotic arguments (the Delta method) or by the bootstrap.

\subsection{Outcome regression estimator} 
\label{sec: outcome regression}

Define $\nu_{a,m}  $ as identification formula \eqref{eq: ident prop 1} and consider the simple regression estimator 
$$\hat{\nu}_{or,a,m}=\frac{1}{n}\sum_{i=1}^{n}\hat{\mathbb{E}}(Y \mid L=L_i, A=a, B=m; \hat{\theta}),$$
where $\mathbb{E}(Y \mid L=l, A=a, B=m; \theta)$ is a parametric model for $\mathbb{E}(Y \mid L=l, A=a, B=m)$ indexed by the parameter ${\theta}$, and assume $ \hat{\theta}$ is its MLE. For example, if $Y$ is binary, we could use a logistic regression model.  The  estimator $\hat{\nu}_{or,a,m}$ is consistent provided that the model indexed by $\theta$ is correctly specified. An analogous parametric g-formula estimator for the expression in Proposition \ref{prop: identification with s}, that also includes $S$, is given in \ref{sec: estimator conditional}. 

\subsection{Weighted estimator}
An inverse probability weighted estimator  $\hat{\nu}_{ipw,a,m} $ of $\nu_{a,m} $ is given by 
$$
\hat{\nu}_{ipw,a,m}=\frac{1}{n}\sum_{i=1}^{n} \frac{I(A_i=a,B_i=m)}{ 
\widehat{\Pr}(B=m \mid L=L_i, A=a; \hat{\alpha}_1) \widehat{\Pr}(A=a \mid L=L_i; \hat{\alpha}_2) }Y_i,
$$
where we have indexed estimated models with the parameter $\alpha_j$ for $j \in \{1, 2\}$ and  assume $ \hat{\alpha}_j$ is its MLE. Often the vaccine assignment probability  $\Pr(A=a \mid L=L_i; \alpha_2) = 0.5$ by design in a RCT, but it is still more efficient to estimate this probability non-parametrically. We derive parametric inverse probability weighted estimators based on the identification result of Proposition \ref{prop: identification with s}, which leverage the additional variable $S$, in  \ref{sec: estimator conditional}.

% given by the solution to the estimating equation $\sum_{i=1}^{n}U_{ipw,i}(\nu_{a,m},\hat{\alpha})=0$ with respect to $\nu_{a,m}$, where
% \begin{align*}
% & U_{ipw,i}(\nu_{a,m},\hat{\alpha}) =    \frac{I(A_i=a,B_i=m)}{ 
% \Pr(B=m \mid L=L_i, A=a; \hat{\alpha}_1) \Pr(A=a \mid L=L_i; \hat{\alpha}_2) } \left(  Y_{i}  - \nu_{a,m} \right), 
% \end{align*}

%\section{Practical implications}
\section{Examples}
\label{sec: examples}

Define vaccine efficacy under interventions on $A$ and $M$ as
\begin{align*}
& VE(m) = 1-  \frac{\mathbb{E} ( Y^{a=1,m})}{  \mathbb{E} ( Y^{a=0,m})}  , \ m \in \{-1,0,1\},
\end{align*}
which is a special case of the generic causal contrasts \eqref{eq: ate} and \eqref{eq: direct eff}. We can interpret  $VE(0)$ and $VE(1)$ as immunological VEs, but the interpretation of $VE(-1)$ as an immunological  VE is more subtle, and depends on whether blinding is broken. Nevertheless, $VE(-1)$ is usually the  implicit target parameter in a blinded RCT.

The  ``total VE'', 
$$
VE_t = 1 - \frac{\mathbb{E} ( Y^{a=1,m=1})}{  \mathbb{E} ( Y^{a=0,m=0})},
$$
is a special case of \eqref{eq: total eff}. 

We consider these parameters in two numerical studies. First, we clarify and compare the values of different vaccine efficacy estimands (Section \ref{sec: VE numerical comparison}). Second, we illustrate the validity of our estimators in simulations based on a real vaccine study  (Section \ref{sec: sims}).   All \textbf{R} scripts used to create the numerical examples are available from \texttt{https://github.com/daniel258/DistinguishingVaccineEffects}.

\subsection{Illustration of numerical differences in vaccine efficacy estimands}
\label{sec: VE numerical comparison}
Consider a hypothetical two-arm vaccine trial $\mathcal{T}_{II}$, where blinding was broken due to a mild side effect of the vaccine. Let the belief $B$ be binary. We are interested in the potential outcomes once the vaccine is available to the population and individuals are fully aware of their vaccination status, so $A=M$. Web Figures \ref{webfig: dag two-arm trial sims} and \ref{webfig: dag six-arm trial sims} present the assumed graphical structures, encoding the key feature that there are no causal or non-causal paths between $M$ and $Y$, except the path $M \rightarrow B \rightarrow Y$.
 
In this scenario, we have data from a trial where placebo recipients are  equally likely to believe that they have received the active vaccine or placebo, $\mathbb{E} (B^{a=0,m=-1})=0.5$. Furthermore,  $\mathbb{E} (Y^{a,m=1})/\mathbb{E} (Y^{a,m=0})=2$ for both $a=0,1$, reflecting an increased risk due to risky behavior when receiving a message $m=1$ compared to $m=0$. Broken blinding is introduced by $\mathbb{E} (B^{a=1,m=-1})/ \mathbb{E} (B^{a=0,m=-1}) = RR_B$, and $RR_B > 1$; treated participants are more likely to believe that they are treated than untreated participants.  Finally, the potential infection rate of unvaccinated people, had they been told they received the vaccine is  $\mathbb{E}(Y^{a=0,m=1})=0.01$. Using these specifications, we can write $VE(0)$ and $VE(1)$ as functions of $VE_t$, and $VE(-1)$ as a function $VE_t$ and $RR_B$, see Appendix \ref{app: sims compare VEs}.

\begin{figure}
    \centering
\includegraphics[scale=0.8]{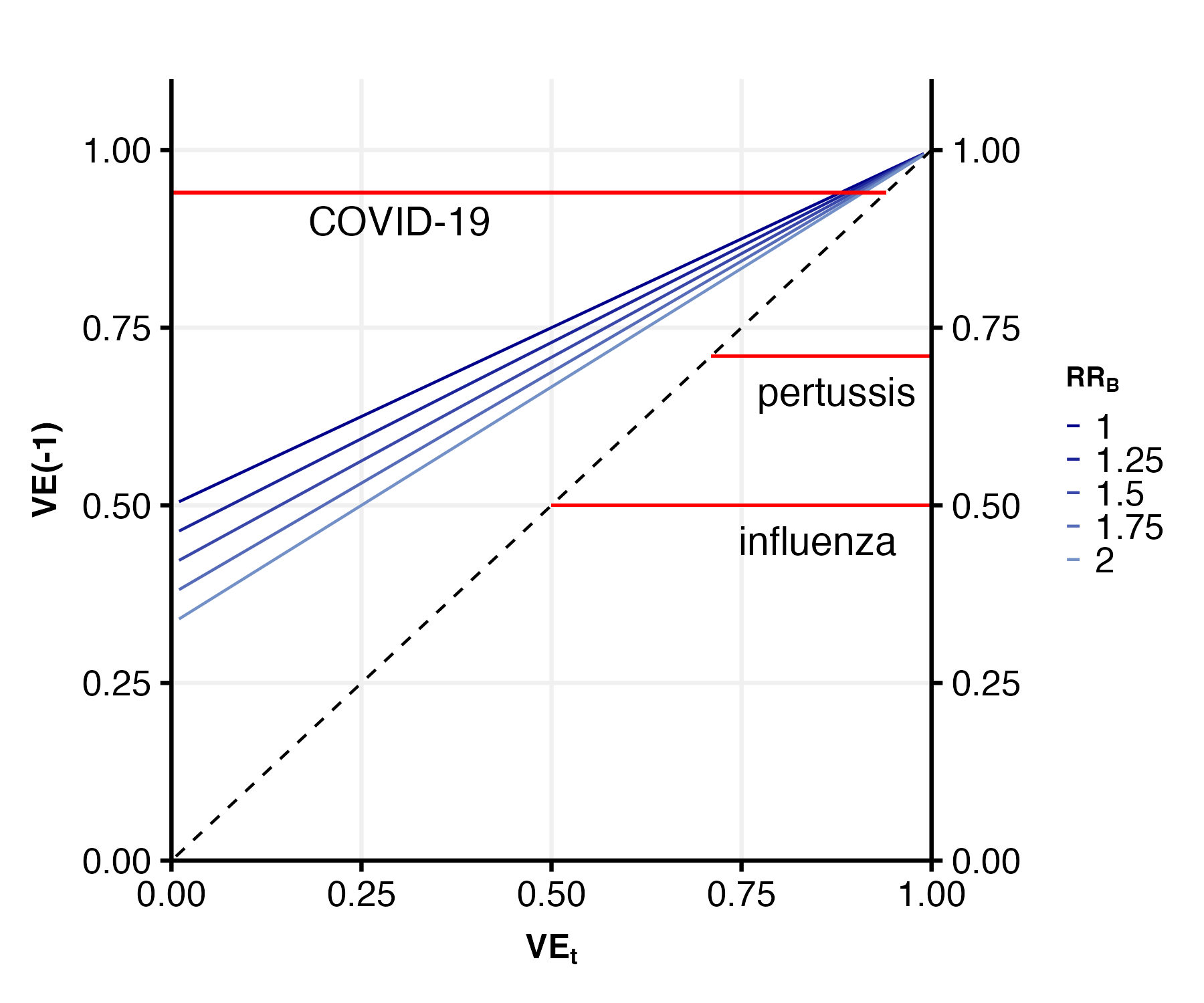}
\caption{$VE(-1)$ versus $VE_t$ as a function of magnitude of broken blinding $RR_B$. Conventional $VE(-1)$ estimates from vaccine RCTs are added as horizontal lines.}
\label{fig: VEminus1 versus total VE}
\end{figure}

Even when blinding is successful, such that $RR_B=1$, $VE(-1)$ might differ from $VE_t$ (Figure \ref{fig: VEminus1 versus total VE}). When $RR_B=1$, then $VE(-1)=VE(0)=VE(1)$. When
 $RR_B>1$, $VE(-1)>VE(m)$ for $m \in \{0,1\}$, and the difference increases as $RR_B$ diverges from one  (Web Figure \ref{webfig: app ve comparisons}). As $RR_B$ increases,   $VE(-1)$ is closer to $VE_t$.
Examples of $VE(-1)$ reported for COVID-19 \citep{polack2020safety}, pertussis \citep{trollfors1995placebo}, and influenza \citep{govaert1994efficacy} are annotated in Figure \ref{fig: VEminus1 versus total VE}.

\subsection{Simulations based on an influenza vaccine trial}
\label{sec: sims}

Here we illustrate that unbiased estimation of $VE_t$ and  $VE(m)$ for $ m \in \{0,1\}$ is possible even when blinding is broken, if side effects $S$ are measured and the conditions of Proposition \ref{prop: identification with s} hold. Our data-generating mechanism (DGM) is grounded in a RCT  \citep{cowling2012protective}, comparing a seasonal influenza vaccine against saline placebo injection in children. The outcome of interest is VE against pandemic influenza A(H1N1) infection, one out of two main outcomes in the trial, coded here as a binary outcome. Our DGM is consistent with the DAG in Web Figure \ref{webfig:vac sb sims} and satisfies the published marginal proportion of adverse reactions in each treatment arm (individual-level data were not published).

%Our outcome of interest was VE against pandemic influenza A(H1N1) infection. We simplified certain aspects of the trial to improve clarity of the illustration, and constructed a DGM consistent with the DAG in Web Figure \ref{webfig:vac sb sims}. 
The original trial was blinded and estimated $VE(-1)$ as 47\%, calculated from the estimated rates $\hat{\mathbb{E}}(Y|A=0)= 0.17$ and $\hat{\mathbb{E}}(Y|A=1)= 0.09$. Furthermore, the trial reported that 50\% of the children in the vaccine arm experienced pain or soreness of any degree at the injection site (41\% mild, 8\% moderate, 1\%) compared to  only 21\% of the children in the placebo arm (19\% mild, 2\% moderate) \citep{cowling2012protective}[Table S1]. 

Let $S^a \in \{0,1\}$ indicate the presence of a side effect under $A=a$. Similar to the trial, let $\mathbb{E}(S^{a=1})= 0.50$ and $\mathbb{E}(S^{a=0})= 0.21$. For $m \in \{0,1\}$, let $B=m$ and for $m=-1$, let $B^{a,s,m}=B^{s}$. Furthermore, $\mathbb{E}(B^{s=1})= 0.70$ and $\mathbb{E}(B^{s=0})= 0.18$, reflecting that those who experience side effects are more likely to believe they received the vaccine. Under these specifications, the magnitude of different vaccine parameters differs substantially (Table \ref{tab:POs sims}). Further technical details about this illustration are given in \ref{app: sims}.
% Setting also the baseline risk $\mathbb{E}(Y^{a=0,m=0})$ and the different $VEs$ (defined as in Section \ref{sec: VE numerical comparison}), Table \ref{tab:POs sims} presents 

We simulated 1,000 datasets from the DGM, with sample sizes corresponding to the trial, 317 in the placebo arm and 479 in the vaccine arm. The observed data vector for each individual consisted of $(A,S,B,Y)$ and $M=-1$. To mitigate finite-sample bias, we repeated the simulations for sample sizes ten times larger (3,170 receiving placebo and 4,790 vaccine).

We estimated  $VE(-1)$ by comparing infection rates for each treatment arm in each dataset, and estimated $\mathbb{E}(Y^{a,m})$ for $a,m \in \{0,1\}$ by substituting expectations with empirical means in \eqref{eq: ident formula S}. We also considered two alternative estimation strategies, where outcomes were compared across treatment arms conditional on $S=s$, for $s \in \{0,1\}$. Such strategies correspond to estimating $$1-\frac{\mathbb{E}(Y^{a=1,m=-1}| S^{a=1,m=-1}=s)}{\mathbb{E}(Y^{a=0,m=-1} | S^{a=0,m=-1}=s )},$$ which generally does not represent a causal effect of interest. Under the assumed DAG (Web Figure \ref{webfig:vac sb sims}), these strategies estimate the controlled direct effect of $A$ on $Y$, while fixing the side effects to be $s$; i.e., when comparing the joint intervention setting $A=1, M=-1, S=s$ versus the intervention $A=0, M=-1, S=s$. 

The results from this DGM indicate that the mean estimates conditional on $S$, $\widehat{VE}(-1, S=0)$ and $\widehat{VE}(-1, S=1)$, are between $VE(m), m=0,1$, and and $VE(-1)$, see Table \ref{tab:POs sims}. In contrast, $VE_t$ is lower than all other VEs. For large sample sizes, our methods gave approximately unbiased estimates. %Therefore, in real-life settings similar to our DGM, the VE estimates obtained naively may overestimate the effect of the vaccine in usual care settings.

\section{Discussion}
% (1)summary (2) how surprising the results are (3) compare and contrast (4) further interpretation and limitations (5) generalizability (imo can be in the limitations) 
%(1)
Contrary to common perceptions, we have argued that the effects targeted in standard vaccine trials often differ from the effects of vaccines in real-world settings. We proposed measuring a single additional variable in a standard trial, analogous to a blinding assessment. Using this variable, and under weak assumptions, it is possible to identify effects that are often relevant to real-world vaccination programs. %Compared to the logistic challenges that usually arise in RCTs, the burden of performing blinding assessments seems to be small.

%(2)

%, our arguments give nuance to  classical VE estimands from RCTs and their practical relevance. However, we also offer a constructive solution, which, to the best of out knowledge, has not been previously implemented in vaccine trials. 
% (3)
To relate our results to previous work in the causal inference literature, we interpret our identification argument in the context of a six arm trial $\mathcal{T}_{VI}$. Our observed data only comprise two out of six arms in $\mathcal{T}_{VI}$, but we target parameters corresponding to expected outcomes in the unobserved four arms. Using this interpretation, $\mathcal{T}_{II}$ assigns a \textit{composite} treatment $(A,M)$, where we only observe individuals with $M=-1$  deterministically. Our identification task is therefore similar to the identification task in classical separable effects settings \citep{robins2010alternative,robins2020interventionist, stensrud2022conditional,stensrud2019separable}. Inspired by the original treatment decomposition idea by \citet{robins2010alternative}, we have decomposed the effect of vaccination into the immunological component of the vaccine, $A$, and a deterministic message, $M=-1$. A similar story of placebo-controlled treatments was given as a motivating example for a treatment decomposition in \citet{didelez2018defining}, who considered interventionist mediation analysis in a survival setting. Our variable definitions are also related to, but different from, the definitions given by Murray \cite{murray2021demystifying}, who explicitly formulated a counterfactual definition of a per-protocol placebo effect (see Web Appendix B).

%also, we can move the "relation to separable effects" here and we will be done with this bit

%(4)
Our proposal has limitations. First, our belief variable requires collection of self-reported data, which may be unreliable. As a remedy, other collected data could be used to assess blinding. As illustrated in our example, the distribution of adverse effects in the two treatment arms could indicate successful blinding. Alternatively, one could perform a negative control outcome analysis \citep{lipsitch2010negative,shi2020selective}. Suppose, for example, that individuals in each arm of an influenza vaccine trial are tested for a panel of other, immunologically distinct respiratory infections when presenting with relevant symptoms. Comparable rates of such other infections would indicate that participants' behavior and exposure patterns are similar across the arms. 

Second, the consequences of forming a belief on behaviour might vary between diseases and populations, due to differences in risk perceptions. Future work should address such  heterogeneity and assess the transportability of vaccine estimates between different settings. 

Third, we defined estimands with respect to a single measurement of belief, but it is possible that beliefs change over time. Similarly, immunological effects are often time-varying, e.g.\ due to waning. In future extensions, we will study longitudinal settings where beliefs, exposures and outcomes vary over time. We conjecture that the methods can be generalized, under appropriate assumptions, by including a time-varying belief variable. 

% Moreover, it may be possible to use a similar technique to leverage observational data to gain insight on the effect of vaccination on behavior...not sure if we want to go into it here 

In conclusion, our arguments give nuance to the practical relevance of  classical VE estimands. But we offer a constructive solution: different estimands, which can quantify immunological and behavioral effects of vaccination in real-world settings, can be identified and estimated under assumptions that often are plausible. %  which, to the best of out knowledge, has not been previously implemented in vaccine trials. 

% A rationale for blinding in vaccine trials is to eliminate behavioral effects of vaccination. However, even under perfect blinding, standard estimands of vaccine efficacy do not correspond to immunological and behavioral effects of vaccination in real-world settings, and we suggest how other estimands can circumvent these issues.

\newpage
\begin{table}[]
    \centering
    \begin{tabular}{c|ccc}
         & $\mathbb{E}(Y^{1,m})$ & $\mathbb{E}(Y^{0,m})$  & $VE(m)$ \\
         \hline
         $m=-1$ & 0.170 & 0.090  & 0.470\\
    $m=0$ &   0.140  & 0.084 & 0.400 \\
    $m=1$ & 0.244 & 0.098 & 0.600
    \end{tabular}
      \caption{Values of $\mathbb{E}(Y^{a,m})$ and $VE(m)$ vaccine efficacy in the simulations modelled after an influenza vaccine RCT. Under this DGM, $VE_t$ equals to $1-{\mathbb{E}(Y^{a=1,m=1})}/{\mathbb{E}(Y^{a=0,m=0})}=0.3$.}
    \label{tab:POs sims}
\end{table}

\begin{table}[]
    \centering
    \small
    \begin{tabular}{cc|cccccc}
    \hline
    &&\multicolumn{6}{c}{Estimand}\\
     & Sample size   & $VE(-1)$ & $VE(0)$ & $VE(1)$ & $VE_t$ & $\widehat{VE}(-1, S=0)$ & $\widehat{VE}(-1, S=1)$\\
    \hline
    True value & & 0.47 & 0.40 & 0.60 & 0.30 &  & \\
    Mean & 796 & 0.463 & 0.380 & 0.575 & 0.273 & 0.446 & 0.526  \\      
     & 7960 & 0.470 & 0.398 & 0.597 & 0.294 & 0.456&  0.555
    \end{tabular}
    \caption{Simulation results for different estimands.}
    \label{tab:estimates sims}
\end{table}

\begin{figure}   
\begin{minipage}{1\textwidth}
        \centering
                \begin{tikzpicture}
                    \tikzset{line width=1.5pt, outer sep=0pt,
                    ell/.style={draw,fill=white, inner sep=2pt,
                    line width=1.5pt},
                    swig vsplit={gap=5pt,
                    inner line width right=0.5pt}};
                    \node[name=A,ell,  shape=ellipse] at (-6,-1) {$A$};
                    \node[name=M,ell,  shape=ellipse] at (-6,1) {$M=-1$};
                    \node[name=B,ell,  shape=ellipse] at (-3,0) {$B$};
                    \node[name=E,ell,  shape=ellipse] at (0,0) {$E$};
                    \node[name=U,ell,  shape=ellipse] at (0,2) {$U$};
                    \node[name=Y,ell,  shape=ellipse] at (3,0) {$Y$};
                     \begin{scope}[>={Stealth[black]},
                                  every edge/.style={draw=black,very thick}]
                        \path [->] (E) edge (Y);
                        \path [->] (A) edge (E);
                        \path [->] (M) edge (B);
                        \path [->] (B) edge (E);
                        \path [->] (U) edge (E);
                        \path [->] (U) edge (Y);
                        \path [->] (E) edge (Y);
                        \path [->] (A) edge[bend left=-30] (Y);    
                    \end{scope}
                \end{tikzpicture}
\subcaption{\label{fig: dag two-arm trial} DAG describing a two-arm trial $\mathcal{T}_{II}$, where $M$ is deterministically equal to $-1$. Thus, the node $M=-1$ is a trivial constant, and is only included in the graph for clarity.}
\end{minipage}
\begin{minipage}{1\textwidth}
        \centering
                \begin{tikzpicture}
                    \tikzset{line width=1.5pt, outer sep=0pt,
                    ell/.style={draw,fill=white, inner sep=2pt,
                    line width=1.5pt},
                    swig vsplit={gap=5pt,
                    inner line width right=0.5pt}};
                    \node[name=A,ell,  shape=ellipse] at (-6,-1) {$A$};
                    \node[name=M,ell,  shape=ellipse] at (-6,1) {$M$};
                    \node[name=B,ell,  shape=ellipse] at (-3,0) {$B$};
                    \node[name=E,ell,  shape=ellipse] at (0,0) {$E$};
                    \node[name=U,ell,  shape=ellipse] at (0,2) {$U$};
                    \node[name=Y,ell,  shape=ellipse] at (3,0) {$Y$};
                     \begin{scope}[>={Stealth[black]},
                                  every edge/.style={draw=black,very thick}]
                        \path [->] (E) edge (Y);
                        \path [->] (A) edge (E);
                        \path [->] (M) edge[line width=0.85mm] (B);
                        \path [->] (B) edge (E);
                        \path [->] (U) edge (E);
                        \path [->] (U) edge (Y);
                        \path [->] (E) edge (Y);
                        \path [->] (A) edge[bend left=-30] (Y);    
                    \end{scope}
                \end{tikzpicture}
\subcaption{\label{fig: dag four-arm trial} DAG describing the (hypothetical) four-arm trial $\mathcal{T}_{IV}$, where the bold arrow indicates the assumed determinism between the message $M$ and the belief $B$.}
\end{minipage}
\begin{minipage}{1\textwidth}
        \centering
                \begin{tikzpicture}
                    \tikzset{line width=1.5pt, outer sep=0pt,
                    ell/.style={draw,fill=white, inner sep=2pt,
                    line width=1.5pt},
                    swig vsplit={gap=5pt,
                    inner line width right=0.5pt}};
                    \node[name=A,ell,  shape=ellipse] at (-6,-1) {$A$};
                    \node[name=M,ell,  shape=ellipse] at (-6,1) {$M$};
                    \node[name=B,ell,  shape=ellipse] at (-3,0) {$B$};
                    \node[name=E,ell,  shape=ellipse] at (0,0) {$E$};
                    \node[name=U,ell,  shape=ellipse] at (0,2) {$U$};
                    \node[name=Y,ell,  shape=ellipse] at (3,0) {$Y$};
                     \begin{scope}[>={Stealth[black]},
                                  every edge/.style={draw=black,very thick}]
                        \path [->] (E) edge (Y);
                        \path [->] (A) edge (E);
                        \path [->] (M) edge (B);
                        \path [->] (B) edge (E);
                        \path [->] (U) edge (E);
                        \path [->] (U) edge (Y);
                        \path [->] (E) edge (Y);
                        \path [->] (A) edge[bend left=-30] (Y);    
                    \end{scope}
                \end{tikzpicture}
\subcaption{\label{fig: dag six-arm trial} DAG describing the hypothetical six-arm trial $\mathcal{T}_{VI}$, where $A \in  \{0,1\}$ and $M \in \{ -1,0,1 \}$ are randomly assigned.}
\end{minipage}
    \caption{DAGs describing $\mathcal{T}_{II}$, $\mathcal{T}_{IV}$ and $\mathcal{T}_{VI}$.}
    \label{fig:dags corresponding to indepedencies]}
\end{figure}
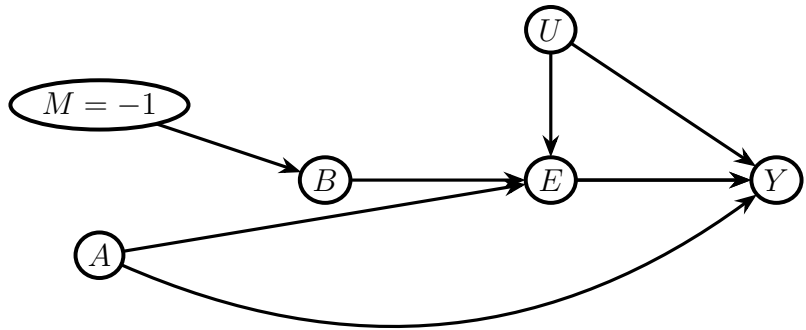
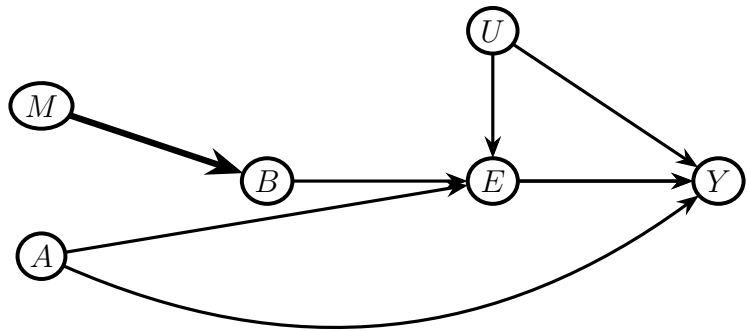
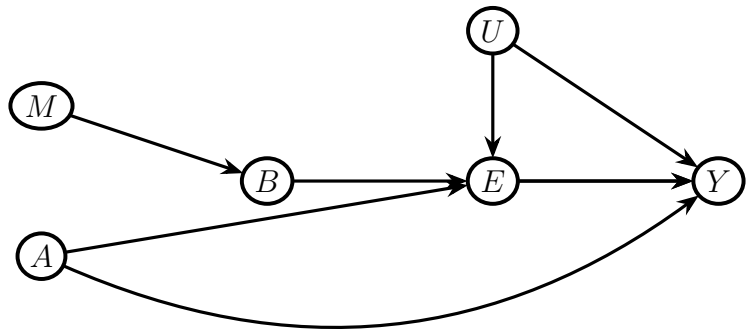

\clearpage

\bibliography{references}
\bibliographystyle{abbrvnat}

\clearpage

\appendix

% Adjust section titles, theorem-, page-, table-, figure- and equation numbers
\renewcommand{\thesection}{Web Appendix \Alph{section}}
\renewcommand{\thesubsection}{\Alph{section}.\arabic{subsection}}
\renewcommand{\thesubsubsection}{\Alph{section}.\arabic{subsection}.\arabic{subsubsection}}

\renewcommand{\figurename}{Web Figure}
\renewcommand{\tablename}{Web Table}

\renewcommand{\theassumption}{\Alph{section}\arabic{assumption}}
\renewcommand{\theproposition}{\Alph{section}\arabic{proposition}}

\renewcommand{\theequation}{\Alph{section}\arabic{equation}}

\section{}
\label{sec: no interference}

\vspace{-0.5cm}

\section*{The no-interference assumption}

Assumptions about no interference between individuals requires careful arguments in vaccine settings, but are often invoked and might sometimes hold. Consider for example vaccines against west Nile virus (WNV), which are currently being developed \citep{gould2023combating}. The no-interference assumption is plausible because WNV is transmitted from birds to humans, with humans having a viral load too low to carry on transmission, constituting dead-end hosts. Thus the infection of one individual, or their vaccination, should not affect the risk of infection in other individuals.

Furthermore, even when the no-interference assumption does not hold exactly in the population, this assumption might hold in the trial. Because nobody outside of the trial receives the vaccine, the only potential source of interference is among trials participants, and the trials are conducted on a small number of individuals relative to the entire population. Even the COVID-19 mRNA vaccine trials, which were large compared to most vaccine trials \citep{polack2020safety, baden2021efficacy}, only included a small proportion of the entire population. Indeed, most analyses of vaccine RCTs rely on the no-interference assumption \citep{tsiatis2021estimating,halloran1996estimability}.

Consider a trial of a new vaccine and let $Y_i^{a_1,...,a_n, a_{n+1},...,a_N}$ be the potential outcome of individual $i \in \mathcal{S}_{
\text{trial}}=\{1,...,n\}$, where $\mathcal{S}_{\text{trial}}$ is the trial sample of size $n$, and $N$ is the population size. Even if the no-interference assumption does not generally hold for a specific triad of a population, disease and vaccine, it can be a reasonable assumption in a trial, as well as in the population when the proportion of vaccinated is still low. Under this particular no interference assumption, 
$$
Y_i^{a_1,...,a_n, a_{n+1}=0,...,a_N=0}=Y_i^{a_i, a_{n+1}=0,...,a_N=0}
$$ 
for $i \in \mathcal{S}_{\text{trial}}$, i.e., vaccination status of other trial participants $j\in \mathcal{S}_{\text{trial}}, j \ne i$ is negligible.
Thus, under the assumption that there is no interference in the trial, the contrast studied in the trial can be written as the direct effect of the vaccine
\begin{equation}
\label{eq: direct effect trial}
E(Y_i^{a_i=1, a_{n+1}=0,...,a_N=0}) \text{ vs. } E(Y_i^{a_i=0, a_{n+1}=0,...,a_N=0}),    
\end{equation}
where here the expectations are taken over $\mathcal{S}_{\text{trial}}$.

The interference in the population can be mild in some scenarios, such as shortly after a vaccination campaign has started. Then, Equation \eqref{eq: direct effect trial} can approximate well the direct effect in the population; that is, the effect of vaccination versus non-vaccination, while fixing the vaccination status of the rest of the population. We leave more formal considerations for future work. 

Therefore, the results presented in the main text  distinguishing between immunological and behavioral vaccine effects can also apply for studying vaccine effects under interference. When interference is present but is negligible in the trial and in the population in the early vaccination  stages, the implicit estimand  in the analysis of the two-arm trial $\mathcal{T}_{II}$ is \eqref{eq: direct effect trial} with the additional setting $m=-1$. 
%Separating immunological and behavioral vaccine effects under substantial interference would entail adapting the notations, assumptions and results from Sections \ref{sec: prelim} and \ref{sec: causal effects} to the potential outcomes and direct effect estimands described in 
%this Section \ref{sec: causal effects two-arm} in the main text.

\section{}
\label{sec: per protocol murray}
\vspace{-0.5cm}

\section*{Per protocol effects and relation to \cite{murray2021demystifying}}

\cite{murray2021demystifying} introduced a random variable $Z$ denoting treatment assignment and $X$ corresponding to received treatment, both taking values in $\{-1,0,1\}$, denoting a nontherapeutic control, placebo and active treatment, respectively. Then, intention-to-treat and per protocol  effects to quantify the placebo effect were defined as 
\begin{align*}
& \mathbb{E} ( Y^{z=-1}) \text{ vs. }  \mathbb{E} ( Y^{z=0}) ,\\
& \mathbb{E} ( Y^{z=-1,x=-1}) \text{ vs. }  \mathbb{E} ( Y^{z=0,x=0})  .    
\end{align*}
To illustrate a point, consider a setting with full compliance such that the per protocol effect is equal to the intention-to-treat effect. Then, the joint of our treatments $(a=0,m=-1)$ will correspond to $(z=-1)$ and $(a=0,m=0)$ will correspond to $(z=0)$. 

We can also extend our results to the setting with incomplete compliance. Let $R$ will be the treatment that is randomly assigned, which consists of our message $M$  and treatment offer $Z$ of taking treatment. Now let $A$ be the treatment (vaccine) actually taken. When the measured baseline covariates $L$ are sufficient to adjust for common causes of $A$ and $Y$, then our arguments extend to the setting in \cite{murray2021demystifying}. 

\section{}
\label{sec: decompositions}
\vspace{-0.5cm}
\section*{Decomposition of the total effect}

In Section \ref{sec: causal effects}, we discussed various causal estimands, in particular the total effect \eqref{eq: total eff} contrasting $\mathbb{E}(Y^{a=1,m=1})$ and
$\mathbb{E}(Y^{a=0,m=0})$. Here we briefly present decompositions of such estimands, which are algebraically similar to decompositions obtained in mediation analysis \citep{robins1992identifiability, pearl2001direct,vanderweele2015explanation}. 

On the difference scale, 
\begin{align}
\begin{split}
\label{eq: tot effect decomp}
\mathbb{E}(Y^{a=1,m=1}) - \mathbb{E}(Y^{a=0,m=0}) &=  \big[\mathbb{E}(Y^{a=1,m=1}) - \mathbb{E}(Y^{a=1,m=0})\big] +  \big[(\mathbb{E}(Y^{a=1,m=0}) -  \mathbb{E}(Y^{a=0,m=0})\big]\\
 &=  \big[\mathbb{E}(Y^{a=1,m=1}) - \mathbb{E}(Y^{a=0,m=1})\big] +  \big[\mathbb{E}(Y^{a=0,m=1}) -  \mathbb{E}(Y^{a=0,m=0})\big]
\end{split}
\end{align}
so the total effect is the sum of two effects: the first line in \eqref{eq: tot effect decomp} presents the total effect as the immunological effect when $m=0$ and the behavioural effect when $a=1$; the second row decomposes the total effect into the behavioral effect without the vaccine ($A$ is held to $a=0$), and the immunological effect when $m=1$.

Analogous results are obtained on the ratio scale, where the decompositions are products rather than sums. Similar results hold for odds ratios when the outcome is rare\citep{vanderweele2010odds}.

Turning to the VE scale, under which the total VE was defined in Section \ref{sec: examples} as 
$VE_t$,  the immunological effects are $VE(m)$, and the behavioral effects \eqref{eq: indirect eff} is $VE_m(a)=1 -     \frac{\mathbb{E}(Y^{a,m=1})}{\mathbb{E}(Y^{a,m=0})}$. Then, 
\begin{align*}
VE_t &= VE_m(0) + VE(1)  - VE_m(0)VE(1)\\
& =VE_m(1) + VE(0)  - VE_m(1)VE(0).
\end{align*}

\section{}
\label{sec: proofs}

\vspace{-0.5cm}
\section*{Proofs}

\subsection{Proof of Proposition \ref{prop: identification simplest}}

\begin{proof}
For $ m \in \{0,1\}$, consider data from the hypothetical four-arm trial $\mathcal{T}_{IV}$, constituting four of the six-arms in the hypothetical trial $\mathcal{T}_{VI}$. We use subscripts to indicate expectations and distributions with respect to a particular trial. For example, $\mathbb{E}_{IV} ( \cdot)$ is the expected value in the superpopulation of $\mathcal{T}_{IV}$. Furthermore, we assume that all participants are drawn from populations with identical distributions of all baseline and pre-baseline covariates, such that, for example $P_{IV}(L=l) = P_{VI}(L=l) = \Pr(L=l)  $, where distributions without subscripts indicate the observed data from the population referring to  $\mathcal{T}_{II}$. 
\begin{align*}
\mathbb{E}(Y^{a,m })  & = \mathbb{E}_{IV}(Y^{a,m }) = \sum_l \mathbb{E}_{IV}(Y^{a,m} \mid L=l)\Pr(L=l) \text{ (LTOT)} \\
  & = \sum_l \mathbb{E}_{IV}(Y^{a,m} \mid M=m,A=a,L=l)\Pr(L=l)  \text{ (randomization of $A$ and $M$)}
 \\
   & = \sum_l \mathbb{E}_{IV}(Y \mid M=m,A=a,L=l)\Pr(L=l)  \text{ (consistency)}
 \\
& = \sum_l \mathbb{E}_{IV}(Y \mid M=m,B=m,A=a,L=l)\Pr(L=l) \text{ ($B=M$ w.p.1 in $\mathcal{T}_{IV}$)} . 
\end{align*}
It follows from Assumptions \ref{ass: delta 1} and \ref{ass: positivity two-arm} that the previous expression is equal to 
\begin{align*}
   &   \sum_l \mathbb{E}(Y \mid M=-1,B=m,A=a,L=l)\Pr(L=l)  \\
 & = \sum_l \mathbb{E}(Y \mid B=m,A=a,L=l)\Pr(L=l) \text{ (Assumption  \ref{ass: delta 1})},  
\end{align*} 
which is observed in $\mathcal{T}_{II^B}$.
\end{proof}

\subsection{Proof of Proposition \ref{prop: identification with s}}

\begin{proof}
For $ m \in \{0,1\}$, consider data from the hypothetical four-arm trial $\mathcal{T}_{IV}$, which also constitute a subset of the data from the six-arm trial $\mathcal{T}_{VI}$. 
\small
\begin{align*}
&  \mathbb{E}(Y^{a,m }) \\
  & = \mathbb{E}_{IV}(Y^{a,m }) = \sum_{l} \mathbb{E}_{IV}(Y^{a,m} \mid L=l) \Pr(L=l) \text{ (LTOT)} \\
  & = \sum_l \mathbb{E}_{IV}(Y^{a,m} \mid M=m,A=a,L=l)\Pr(L=l)  \text{ ($A,M$ randomized)}
 \\
  & = \sum_l \mathbb{E}_{IV}(Y^{a,m} \mid M=m,A=a,L=l)\Pr(L=l \mid A =a)  \text{ ($A$ randomized)}
 \\
   & = \sum_l \mathbb{E}_{IV}(Y \mid M=m,A=a,L=l)\Pr(L=l  \mid A =a)  \text{ (consistency)}
 \\ %\mid S=s, L=l) \Pr(S=s, L=l)
   & = \sum_{l,s} \mathbb{E}_{IV}(Y \mid M=m,A=a, S=s, L=l) {\Pr}_{IV}(S=s \mid L=l , A =a) {\Pr}_{IV}(L=l)  \\
  & \quad (\eqref{eq: delta m s}, \text{ LTOT, $A$ randomized})
 \\ %
&    = \sum_{l,s} \mathbb{E}_{IV}(Y \mid M=m,B=m,A=a, S=s, L=l) {\Pr}_{IV}(S=s \mid L=l , A =a) {\Pr}_{IV}(L=l) \  \text{($B=M$ w.p.1)} .
\end{align*}
It follows from Assumptions \ref{ass: delta 2} and \ref{ass: positivity side eff} that the previous expression is equal to 
\begin{align*}
   &  = \sum_{l,s} \mathbb{E}(Y \mid M=-1,B=m,A=a,S=s, L=l) \Pr(S=s, \mid L=l , A =a) \Pr(L=l)  \\
  &  = \sum_{l,s} \mathbb{E}(Y \mid B=m,A=a,S=s, L=l) \Pr(S=s, \mid L=l , A =a) \Pr(L=l) ,  
\end{align*} 
where we also used that $A$ is randomized, that ${\Pr}_{IV}(S=s \mid L=l , A =a)= \Pr(S=s \mid L=l , A =a)$ by assumption and that $\Pr(M=-1 ) =1$ in $\mathcal{T}_{II}$,%which observable in the (original) two-arm trial
\end{proof}

\section{Estimation conditional on the presence of a side effect $S$}
\label{sec: estimator conditional}
\subsection{Outcome regression estimator} 
Consider first a simple regression estimator $\hat{\theta}_{or,a,m}$, where we define $\theta_{a,m}  $ as the identification formula in Proposition \ref{prop: identification with s}. Let this estimator be the solution to $\sum_{i=1}^{n}U_{or,i}(\theta_{a,m},\hat{\eta})=0$ with respect to $\theta_{a,m}$, where
\begin{align*}
& U_{or,i}(\theta_{a,m},\hat{\eta}) =  \\
&  \sum_s  \mathbb{E}(Y \mid B=m,A=a,S=s, L_i; \hat{\eta}_1) \Pr(S=s, \mid L_i , A =a; \hat{\eta}_2) - \theta_{a,m}  .
\end{align*}

such that $  \mathbb{E}(Y \mid B=m,A=a,S=s, L=l; \hat{\eta}_1) $ and $ \Pr(S=s, \mid L=l , A =a; \hat{\eta}_2) $ are parametric models indexed by the parameter ${\eta}_k, \ k=1,2$, where $ \hat{\eta}_k$ is its MLE. The estimating equation defined by $U_{or,i}(\theta_{a,m},\hat{\eta})$ has mean zero and the estimator $\hat{\eta}_{or,a,m}$ is consistent provided that the models are correctly specified. 

Alternatively, consider the estimator $\sum_{i=1}^{n}U'_{or,i}(\theta_{a,m},\hat{\eta}_1)=0$ with respect to $\theta_{a,m}$, where 
\begin{align*}
U'_{or,i}(\theta_{a,m},\hat{\eta}_1) =  
& I(A_i=a) \{ \mathbb{E}(Y \mid B = m, A = a, S = S_i, L = L_i; \hat{\eta}_1) - \theta_{a,m} \}  ,
\end{align*}
which is also consistent and requires specification of a smaller number of parametric models, but also uses less data and thus might be less efficient in the setting where all models are correctly specified. 

\subsection{Weighted estimator}
An inverse probability weighted estimator  $\hat{\theta}_{ipw,a,m} $ of $\theta_{a,m} $ is the solution to the estimating equation $\sum_{i=1}^{n}U_{ipw,i}(\theta_{a,m},\hat{\beta})=0$ with respect to $\theta_{a,m}$, where
\begin{align*}
& U_{ipw,i}(\theta_{a,m},\hat{\beta}) =    \frac{I(A_i=a,B_i=m)}{ \Pr(B=m \mid L=L_i, S=S_i, A=a; \hat{\beta}_1 ) \Pr(A=a \mid L=L_i ;\hat{\beta}_2) } \left(  Y_{i}  - \theta_{a,m} \right), 
\end{align*}
where we have indexed estimators with the parameter $ \beta_j$ for $j \in \{1, 2\}$ and assume $ \hat{\beta}_j$ is its MLE. As stated in the main text, usually $\Pr(A=a \mid L= L_i) = 0.5$ by design in a vaccine RCT, but it can also be estimated as $\Pr(A=a \mid L=L_i ;\hat{\beta}_2)$ for efficiency reasons.

\section{}
\label{app: sims}

\vspace{-0.5cm}

\section*{Further details and results for the examples}

\subsection{Illustration of different vaccine effects}
\label{app: sims compare VEs}

The mathematical of the DGM described in Section \ref{sec: VE numerical comparison} of the main text is as follows. As previously noted, this illustration is presented under the graphical structure of Web Figures \ref{webfig: dag two-arm trial sims} and \ref{webfig: dag six-arm trial sims}.
 %We also take  ${\mathbb{E} (Y^{a,m=1})} / {\mathbb{E} (Y^{a,m=0})}=2$ for both $a=0,1$, reflecting an increased risk due to a more risky behavior when receiving a message $m=1$ compared to $m=0$. Broken blinding is introduced by letting $\mathbb{E} (B^{a=1,m=-1})/ \mathbb{E} (B^{a=0,m=-1}) = RR_B$, and setting $RR_B > 1$. 
The potential outcomes of $B$ under interventions on $A,M$ were
\begin{align*}
& B^{a,m=0}=0, \quad\text{for} \quad a=0,1\\
& B^{a,m=1}=1, \quad \text{for} \quad a=0,1\\
& \mathbb{E} (B^{a=0,m=-1}=1) = 0.5 \\
& \mathbb{E} (B^{a=1,m=-1}=1) = 0.5\times RR_{B} \\
\end{align*}
The potential outcomes of $Y$ under interventions on $A,M$ were
\begin{align}
\begin{split}
    \label{Eq: VE comparison POdef}
&\mathbb{E} (Y^{a=0,m=0} ) = 0.01  \\
&\mathbb{E} (Y^{a=1,m=1} ) = 0.01\times(1- VE_t)^{\ddagger}  \\
&\mathbb{E} (Y^{a=0,m=1}) = 0.02^{\star} \\
&\mathbb{E} (Y^{a=1,m=0}) = 0.005\times(1- VE_t)^{\star} \\
\end{split}
\end{align}
 $^\ddagger$ By the definition of $VE_t$.  \\
$^\star$ By  $\frac{\mathbb{E} (Y^{a,m=1})}{\mathbb{E} (Y^{a,m=0})}=2$ for both $a=0,1$

Under the above DGM, we can calculate $VE(m)$. We start with $VE(0)$ and $VE(1)$ which are immediately given by \eqref{Eq: VE comparison POdef} as
\begin{align*}
    VE(0) &= 1 - \frac{\mathbb{E} (Y^{a=1,m=0})}{\mathbb{E} (Y^{a=0,m=0})} = 0.5(VE_t + 1)\\
    VE(1) &= 1 - \frac{\mathbb{E} (Y^{a=1,m=1})}{\mathbb{E} (Y^{a=0,m=1})} = 0.5(VE_t + 1)
\end{align*}

Turning to $VE(-1)$, note that because there are no causal and no non-casual paths between $M$ and $Y$ except the indirect effect through $B$, and because $B^{a,m}=m$ for $a,m=0,1$, we have
 \begin{align*}
      \mathbb{E}(Y^{a=0,m=-1} | B^{a=0,m=-1} = 0 ) &= \mathbb{E}(Y^{a=0,m=0} | B^{a=0,m=0} = 0 )= \mathbb{E}(Y^{a=0,m=0}),\\
      \mathbb{E}(Y^{a=0,m=-1} | B^{a=0,m=-1} = 1 ) &= \mathbb{E}(Y^{a=0,m=1} | B^{a=0,m=1} = 1 )= \mathbb{E}(Y^{a=0,m=1}),\\
      \mathbb{E}(Y^{a=1,m=-1} | B^{a=1,m=-1} = 1 ) &= \mathbb{E}(Y^{a=1,m=1} | B^{a=0,m=1} = 1 )= \mathbb{E}(Y^{a=1,m=1}),\\
       \mathbb{E} (Y^{a=1,m=-1} | B^{a=1,m=-1} = 0) &= \mathbb{E} (Y^{a=1,m=0} | B^{a=1,m=0} = 0) = \mathbb{E} (Y^{a=1,m=0}).       
 \end{align*}
 We can now use the above results to derive $\mathbb{E} (Y^{a=0,m=-1})$ and $\mathbb{E} (Y^{a=1,m=-1})$, as follows.
\begin{align*}
    \mathbb{E} &(Y^{a=0,m=-1}) \\[1ex]
    &= \mathbb{E} (Y^{a=0,m=-1}  | B^{a=0,m=-1} = 1)\times0.5 + \mathbb{E} (Y^{a=0,m=-1}  | B^{a=0,m=-1} = 0)\times0.5 \\[1ex]
    &= \mathbb{E} (Y^{a=0,m=1})\times0.5 +  \mathbb{E} (Y^{a=0,m=0})\times0.5 \\[1ex]
    &=  0.03\times0.5 = 0.015.
\end{align*}
Next,
\begin{align*}
    \mathbb{E} &(Y^{a=1,m=-1}) \\[1ex]
    &= \mathbb{E} (Y^{a=1,m=-1}  | B^{a=1,m=-1} = 1)\times0.5\times RR_B + \mathbb{E} (Y^{a=1,m=-1}  | B^{a=1,m=-1} = 0)\times(1 - 0.5RR_B) \\[1ex]
    &= \mathbb{E} (Y^{a=1,m=1})\times0.5\times RR_B +  \mathbb{E} (Y^{a=1,m=0})\times(1 - 0.5\times RR_B) \\[1ex]
    &=  0.01\times(1- VE_t)\times0.5\times RR_B +  0.005\times(1- VE_t) \times(1 - 0.5\times RR_B) \\
    &= (1- VE_t)\times(0.005+ 0.0025\times RR_B).
    \end{align*}

Therefore, 
$$
VE(-1) = 1 - \frac{(1- VE_t)\times(0.005 + 0.0025\times RR_B)}{0.015}
% VE(-1) = 1 - \frac{(1- VE_t)\times(0.007 + 0.015\times RR_B)}{0.0075}
$$

\pagebreak
\begin{figure}   
\begin{minipage}{1\textwidth}
        \centering
                \begin{tikzpicture}
                    \tikzset{line width=1.5pt, outer sep=0pt,
                    ell/.style={draw,fill=white, inner sep=2pt,
                    line width=1.5pt},
                    swig vsplit={gap=5pt,
                    inner line width right=0.5pt}};
                    \node[name=A,ell,  shape=ellipse] at (-6,-1) {$A$};
                    \node[name=M,ell,  shape=ellipse] at (-6,1) {$M=-1$};
                    \node[name=B,ell,  shape=ellipse] at (-3,0) {$B$};
                    %\node[name=E,ell,  shape=ellipse] at (0,0) {$E$};
                    %\node[name=U,ell,  shape=ellipse] at (0,2) {$U$};
                    \node[name=Y,ell,  shape=ellipse] at (0,0) {$Y$};
                     \begin{scope}[>={Stealth[black]},
                                  every edge/.style={draw=black,very thick}]
                 %       \path [->] (E) edge (Y);
                     %   \path [->] (A) edge (E);
                        \path [->] (A) edge (B);
                        \path [->] (M) edge (B);
                        \path [->] (B) edge (Y);
                        %\path [->] (U) edge (E);
                        %\path [->] (U) edge (Y);
                        %\path [->] (E) edge (Y);
                        \path [->] (A) edge[bend left=-30] (Y);    
                    \end{scope}
                \end{tikzpicture}
\subcaption{\label{webfig: dag two-arm trial sims} DAG describing the two-arm trial $\mathcal{T}_{II}$ from Section \ref{sec: VE numerical comparison}, where $M$ is deterministically equal to $-1$. Thus, the node $M=-1$ is a trivial constant, and is only included for clarity.}
\end{minipage}
\begin{minipage}{1\textwidth}
        \centering
                \begin{tikzpicture}
                    \tikzset{line width=1.5pt, outer sep=0pt,
                    ell/.style={draw,fill=white, inner sep=2pt,
                    line width=1.5pt},
                    swig vsplit={gap=5pt,
                    inner line width right=0.5pt}};
                    \node[name=A,ell,  shape=ellipse] at (-6,-1) {$A$};
                    \node[name=M,ell,  shape=ellipse] at (-6,1) {$M$};
                    \node[name=B,ell,  shape=ellipse] at (-3,0) {$B$};
                 %   \node[name=E,ell,  shape=ellipse] at (0,0) {$E$};
                  %  \node[name=U,ell,  shape=ellipse] at (0,2) {$U$};
                    \node[name=Y,ell,  shape=ellipse] at (0,0) {$Y$};
                     \begin{scope}[>={Stealth[black]},
                                  every edge/.style={draw=black,very thick}]
                    %    \path [->] (E) edge (Y);
                   %     \path [->] (A) edge (E);
                        \path [->] (A) edge (B);
                        \path [->] (M) edge (B);
                        \path [->] (B) edge (Y);
                 %       \path [->] (U) edge (E);
                  %      \path [->] (U) edge (Y);
                       % \path [->] (E) edge (Y);
                        \path [->] (A) edge[bend left=-30] (Y);    
                    \end{scope}
                \end{tikzpicture}
\subcaption{\label{webfig: dag six-arm trial sims} DAG describing the six-arm trial $\mathcal{T}_{VI}$ from Section \ref{sec: VE numerical comparison}. The arrows from $M$ and $A$ to $B$ reflect that when $M =0,1$, then $B=M$, and when $M=-1$, then $B$ is a random variable whose distribution depends on $A$.  }
\end{minipage}
\caption{Figures describing the DGM used in Section \ref{sec: VE numerical comparison}.}
\end{figure}

\begin{figure}
    \centering
\includegraphics[scale=0.5]{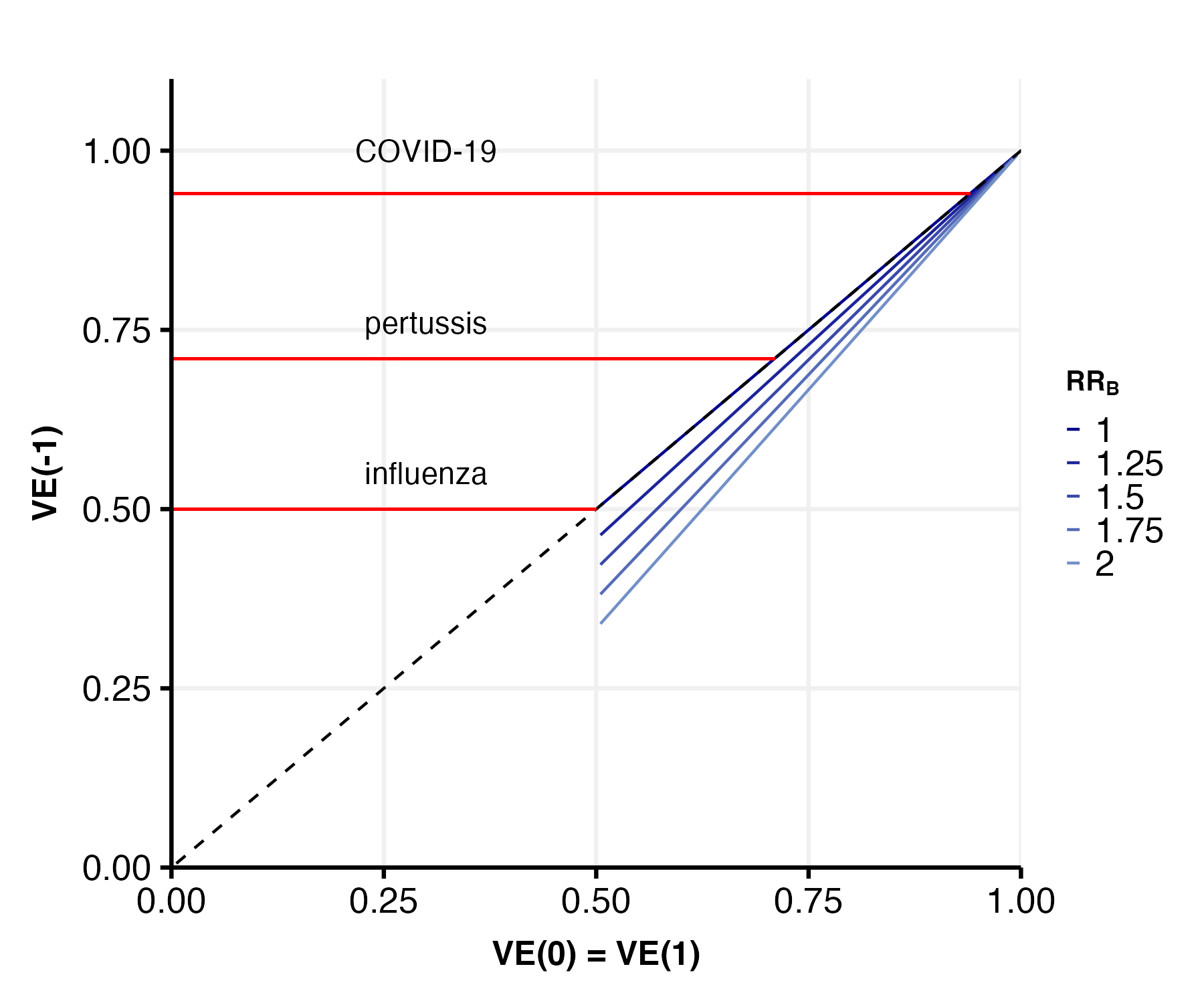}
\includegraphics[scale=0.5]{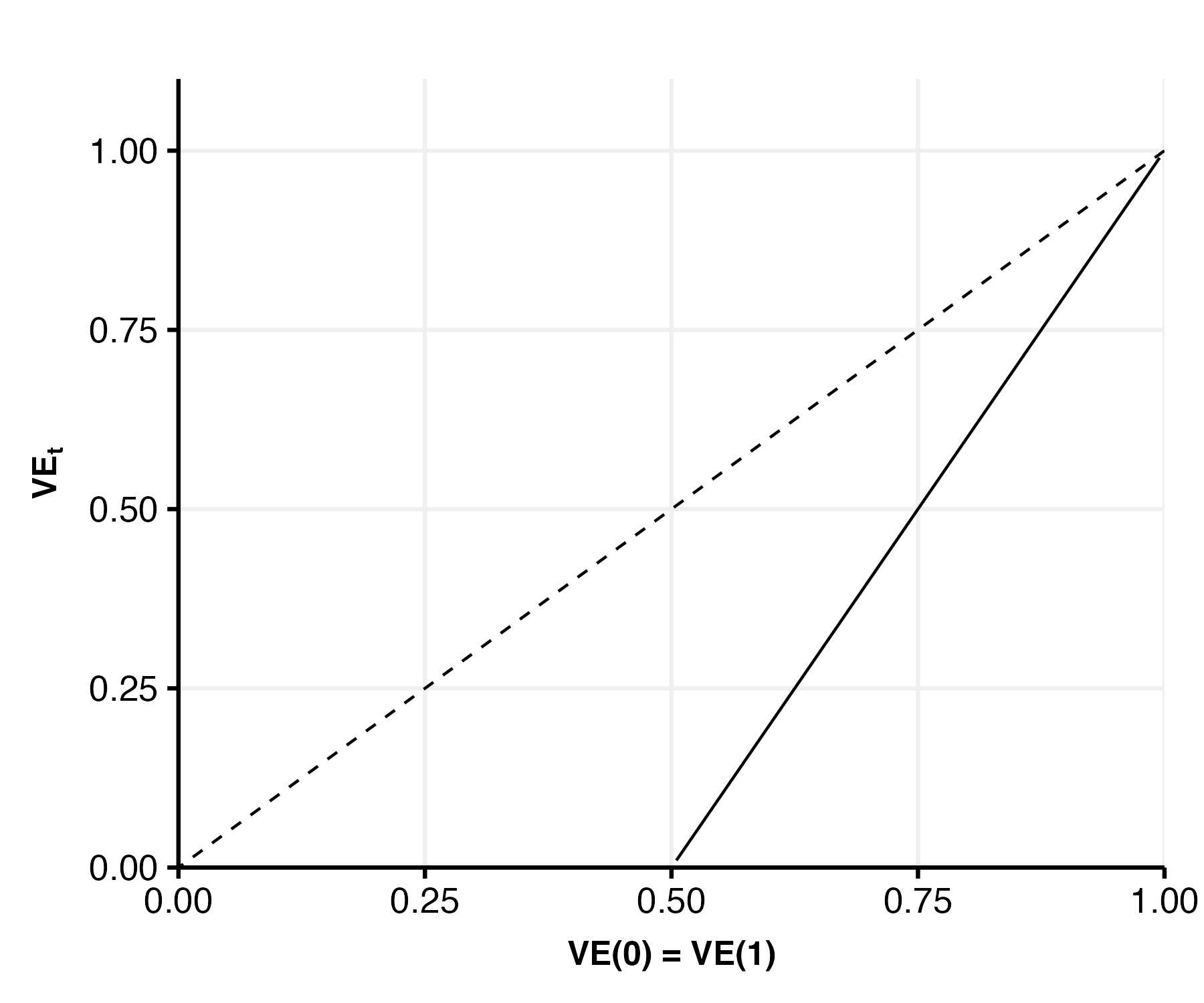}
    \caption{$VE(-1)$ and $VE_t$ as a function of $VE(m), m=0,1$ for different $RR_B$ values. Note that $VE(m), m=0,1$ and $VE_t$ are not functions of $RR_B$, because the belief is identical to the message in these comparisons.}
    \label{webfig: app ve comparisons}
\end{figure}

Web Figure     \ref{webfig: app ve comparisons} presents an additional comparison, between $VE(-1)$ as a function of $VE(0)=VE(1)$ for different $RR_B$ values and the relationship between $VE_t$ and $VE(0)=VE(1)$. Because it is assumed a message $M=1$ induces more risky behaviour, we obtained that $VE_t < VE(m), m=0,1$.

\subsection{Simulations}
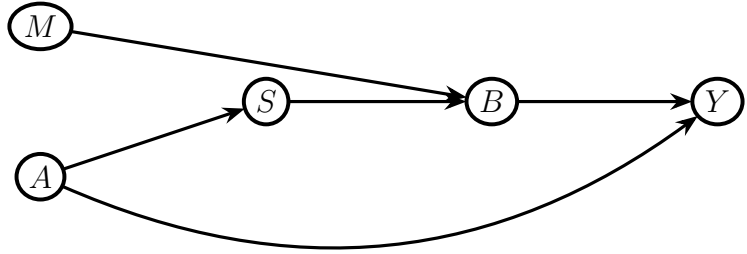
\begin{figure}
    \begin{minipage}{1\textwidth}
        \centering
                \begin{tikzpicture}
                    \tikzset{line width=1.5pt, outer sep=0pt,
                    ell/.style={draw,fill=white, inner sep=2pt,
                    line width=1.5pt},
                    swig vsplit={gap=5pt,
                    inner line width right=0.5pt}};
                    \node[name=A,ell,  shape=ellipse] at (-9,-1) {$A$};
                    \node[name=S,ell,  shape=ellipse] at (-6,0) {$S$};
                    \node[name=B,ell,  shape=ellipse] at (-3,0) {$B$};
                    %\node[name=E,ell,  shape=ellipse] at (0,0) {$E$};
              %      \node[name=U,ell,  shape=ellipse] at (-1.5,2) {$U$};
                    \node[name=Y,ell,  shape=ellipse] at (0,0) {$Y$};
                    \node[name=M,ell,  shape=ellipse] at (-9,1) {$M$};
                     \begin{scope}[>={Stealth[black]},
                                  every edge/.style={draw=black,very thick}]
                        %\path [->] (E) edge (Y);
                        \path [->] (A) edge (S);
                        \path [->] (S) edge (B);
                        \path [->] (M) edge (B);
               %         \path [->] (A) edge[line width=0.85mm] (M);
                        \path [->] (B) edge (Y);
                        %\path [->] (U) edge (E);
               %         \path [->] (U) edge[color=orange] (S);
                %        \path [->] (U) edge (Y);
                        %\path [->] (E) edge (Y);
                        %\path [->] (S) edge[bend left=-30] (Y);    
                        \path [->] (A) edge[bend left=-30] (Y);    
                    \end{scope}
                \end{tikzpicture}
\end{minipage}
\caption{\label{webfig:vac sb sims}DAG describing the simulations according to a six-arm trial  $\mathcal{T}_{VI}$, where a side effect $S$ can affect the belief $B$.}
\end{figure}

Web Figure \ref{webfig:vac sb sims} describes the assumptions underlying the DGM used in our simulations.
Starting from the non-trial conditions ($A=M=B$), $\mathbb{E} (Y^{a,m})$ are determined by setting $\mathbb{E} (Y^{a=0,m=0})$, $VE_{t}$, and $VE(m)$ for $m=0,1$. The values will be determined later to obtain similar results to the trial. Next, to calculate $\mathbb{E} (Y^{a,m=-1})$, we note that for $m=-1$, only $S$ affects $B$, and we can write
% \begin{align*}
% \mathbb{E} &(Y^{a,m=-1})\\ 
% =  \mathbb{E} &(Y^{a,m=-1} | B^{s=1,m=-1}=1,S^{a=1}=1)\Pr(S^{a}=0)\Pr(B^{s=0,m=-1}=1)\\[1 ex]
% &+\mathbb{E} (Y^{a,m=-1} | B^{s=0,m=-1}=1,S^{a=1}=1)\Pr(S^{a}=1)\Pr(B^{s=1,m=-1}=1)\\[1 ex]
% &+\mathbb{E} (Y^{a,m=-1} | B^{s=1,m=-1}=0,S^{a=1}=0)\Pr(S^{a}=0)\Pr(B^{s=0,m=-1}=0)\\[1 ex]
% &+\mathbb{E} (Y^{a,m=-1} | B^{s=1,m=-1}=0,S^{a=1}=1)\Pr(S^{a}=1)\Pr(B^{s=1,m=-1}=0)\\[2 ex]
%  =  \mathbb{E} &(Y^{a,m=1})\big[\Pr(S^{a}=0)\Pr(B^{s=0}=1)+ \Pr(S^{a}=1)\Pr(B^{s=1}=1)\big]\\[1 ex]
% &+\mathbb{E} (Y^{a,m=0})\big[ \Pr(S^{a}=0)\Pr(B^{s=0}=0)
% +\Pr(S^{a}=1)\Pr(B^{s=1}=0)\big]\\[1 ex]
% \end{align*}

\begin{align*}
\mathbb{E} &(Y^{a,m=-1})\\ 
=  \mathbb{E} &(Y^{a,m=-1} | B^{s=0,m=-1}=1,S^{a=1}=0)\Pr(S^{a}=0)\Pr(B^{s=0,m=-1}=1)\\[1 ex]
&+\mathbb{E} (Y^{a,m=-1} | B^{s=1,m=-1}=1,S^{a=1}=1)\Pr(S^{a}=1)\Pr(B^{s=1,m=-1}=1)\\[1 ex]
&+\mathbb{E} (Y^{a,m=-1} | B^{s=0,m=-1}=0,S^{a=1}=0)\Pr(S^{a}=0)\Pr(B^{s=0,m=-1}=0)\\[1 ex]
&+\mathbb{E} (Y^{a,m=-1} | B^{s=1,m=-1}=0,S^{a=1}=1)\Pr(S^{a}=1)\Pr(B^{s=1,m=-1}=0)\\[2 ex]
 =  \mathbb{E} &(Y^{a,m=1})\big[\Pr(S^{a}=0)\Pr(B^{s=0}=1)+ \Pr(S^{a}=1)\Pr(B^{s=1}=1)\big]\\[1 ex]
&+\mathbb{E} (Y^{a,m=0})\big[ \Pr(S^{a}=0)\Pr(B^{s=0}=0)
+\Pr(S^{a}=1)\Pr(B^{s=1}=0)\big]\\[1 ex]
\end{align*}

Note these weights can be abbreviated a bit recognizing it's just the probability of belief
% \mathbb{E} (Y^{a,m=-1}) &=  \mathbb{E} (Y^{a,m=-1} | B^{1,m=-1}=1,S^{a=1}=1)\Pr(S^{a=1})\Pr(B^{s=1,m=-1}=1)\\
Now, under the values given below, we obtain population-level results similar to what was obtained in the trial. We took $\mathbb{E} (Y^{a,m=-1})$ (and therefore $VE(-1)$) to be approximately equal to $\hat{\Pr}(Y=1|A=a)$ from the trial. The estimated VE in the trial was 0.47. We also took the probability of side effects under each treatment $\Pr(S^{a}=1)$ to be equal to the trial observed proportions of soreness of any severity at each treatment  arm.
We took
\begin{align*}
\mathbb{E}(Y^{a=0,m=-1}) &= 0.170 \\
\mathbb{E}(Y^{a=1,m=-1}) &= 0.090 \\
    \mathbb{E}(Y^{a=0,m=0}) &= 0.1395 \\
    VE_{tot} &= 0.3\\
    VE(0) &= 0.40\\
    VE(1) &= 0.60\\
    \Pr(S^{a=1}=1) &= 0.50 \\
    \Pr(S^{a=0}=1) &= 0.21\\
    \Pr(B^{s=1}=1) &= 0.70\\
    \Pr(B^{s=0}=1) &= 0.18\\
\end{align*}
which resulted in the mean potential outcomes $\mathbb{E}(Y^{a,m})$ given in Table \ref{tab:POs sims} of the main text.

\end{document}

\def\spacingset#1{\renewcommand{\baselinestretch}%
{#1}\small\normalsize} \spacingset{1}

%%%%%%%%%%%%%%%%%%%%%%%%%%%%%%%%%%%%%%%%%%%%%%%%%%%%%%%%%%%%%%%%%%%%%%%%%%%%%%

\if1\blind
{
%  \title{\bf Switching treatment using an early treatment response on recurrent outcomes}
   \title{\bf Web Appendix for ``Distinguishing immunological and behavioral effects of vaccination''}
  \author{Authors
    \\
    Department of Mathematics, EPFL}
  \maketitle
} \fi

\if0\blind
{
  \bigskip
  \bigskip
  \bigskip
  \begin{center}
    {\LARGE\bf }
\end{center}
  \medskip
} \fi

% \bigskip

\spacingset{1.9} % DON'T change the spacing!

\allowdisplaybreaks % Allow page breaks in align environment

\appendix

% Adjust section titles, theorem-, page-, table-, figure- and equation numbers
\renewcommand{\thesection}{Web Appendix \Alph{section}}
\renewcommand{\thesubsection}{\Alph{section}.\arabic{subsection}}
\renewcommand{\thesubsubsection}{\Alph{section}.\arabic{subsection}.\arabic{subsubsection}}

\renewcommand{\figurename}{Web Figure}
\renewcommand{\tablename}{Web Table}

\renewcommand{\theassumption}{\Alph{section}\arabic{assumption}}
\renewcommand{\theproposition}{\Alph{section}\arabic{proposition}}

\renewcommand{\theequation}{\Alph{section}\arabic{equation}}